\documentclass[%
 reprint,
superscriptaddress,
 amsmath,amssymb,
 aps,
 prf,
]{revtex4-2}
\newcommand{\um}{\text{\textmu{}m}}
\newcommand{\etal}{\textit{et~al.}}
\usepackage{textcomp}
\usepackage{units}
\usepackage{hyperref}
\usepackage{graphicx}
\usepackage{dcolumn}
\usepackage{bm}

\begin{document}

\preprint{APS/123-QED}

\title{Cell-free layer of red blood cells in a constricted microfluidic channel under steady and time-dependent flow conditions}

\author{Steffen M. Recktenwald}
 \email{steffen.recktenwald@uni-saarland.de}
 \affiliation{Dynamics of Fluids, Department of Experimental Physics, Saarland University, Saarbr{\"u}cken, Germany}
\author{Katharina Graessel}%
\affiliation{Biofluid Simulation and Modeling, Department of Physics, University of Bayreuth, Bayreuth, Germany}%
\author{Yazdan Rashidi}
 \affiliation{Dynamics of Fluids, Department of Experimental Physics, Saarland University, Saarbr{\"u}cken, Germany}%
\author{Jann Niklas Steuer}
 \affiliation{Dynamics of Fluids, Department of Experimental Physics, Saarland University, Saarbr{\"u}cken, Germany}%
 \author{Thomas John}
 \affiliation{Dynamics of Fluids, Department of Experimental Physics, Saarland University, Saarbr{\"u}cken, Germany}%
 \author{Stephan Gekle}
 \affiliation{Biofluid Simulation and Modeling, Department of Physics, University of Bayreuth, Bayreuth, Germany}%
 \author{Christian Wagner}
 \affiliation{Dynamics of Fluids, Department of Experimental Physics, Saarland University, Saarbr{\"u}cken, Germany}%
 \affiliation{Physics and Materials Science Research Unit, University of Luxembourg, Luxembourg, Luxembourg}
\date{\today}

\begin{abstract}
Constricted blood vessels in the circulatory system can severely impact the spatiotemporal organization of red blood cells (RBCs) causing various physiological complications. In lab-on-a-chip applications, constrictions are commonly used for cell sorting and plasma separation based on the formation of a cell-free layer (CFL) at the channel boundary. However, such devices usually employ a steady flow, although time-dependent and pulsatile flow conditions enhance many microfluidic operations and possess significant relevance for in-vitro studies under physiologically relevant flow conditions. In this study, we examine the CFL dynamics in a constricted microchannel under steady and time-dependent flow. Therefore, we develop an image processing routine that allows us to resolve the spatiotemporal evolution of the CFL under time-dependent driving of the flow with arbitrary waveform. First, we perform a characterization of the CFL and the cell-free area (CFA) before and after the constriction under steady flow conditions. Second, we employ our method to study the effect of the hematocrit and the parameters of the flow modulations on the CFL and CFA using a sinusoidal pressure profile. Our results highlight the dominant effects of the RBC concentration and the amplitude of the applied pressure signal predominantly on the CFA dynamics. Moreover, we observe a dampening of the CFA amplitude with increasing hematocrit or decreasing pressure amplitude, and a peculiar phase shift of the CFA oscillations pre- and post-constriction. Complementary numerical simulations reveal how the time-dependent CFA dynamics are coupled with the dynamically changing flow field at the constriction. Due to the increasing demand for investigations and applications under time-dependent flow conditions, our study provides crucial insight into the flow behavior of complex fluids in unsteady microscale flows.
\end{abstract}

\maketitle

\section{Introduction}
The circulatory system is a complex vessel network that distributes blood to body tissues and organs. Red blood cells (RBCs) are the main constituent of human blood and occupy up to \mbox{$\unit[45]{\%}$} of the blood volume. In the microcirculation, RBCs migrate away from the vessel walls, thus forming a core RBC flow and a cell-free layer (CFL). This impacts the lateral movement and spatiotemporal distribution of RBCs and crucially affects blood rheology by reducing the apparent viscosity of blood in small vessels.\cite{Fahraeus1929a, Fahraeus1931} The formation of the CFL near the vessel walls determines the unique flow properties of blood in-vivo,\cite{Pries2008, Secomb2017} impacting various physiological phenomena, such as platelet and leukocyte margination,\cite{Freund2007a, Zhao2012} the wall shear stress sensed by the endothelial cells,\cite{Davies1993} and the heterogeneous RBC distribution in branching vessels and networks.\cite{Bento2018, Balogh2019, Bento2019} Additionally, RBC organization and CFL formation play a pivotal role in biomedical lab-on-a-chip applications for cell and plasma separation and in flow-focusing microfluidic devices.\cite{Yang2006, Sollier2010, Lee2011a, Lee2011b, Pinho2013, Tripathi2015} Therefore, it is paramount to understand how the local and temporal RBC distribution affects blood flow and hemorheology.

The microscale collective behavior of RBCs, such as their organization and the emergence of a CFL close to the vessel wall, has been extensively studied experimentally,\cite{Bugliarello1970, Fenton1985, Lima2008, Bento2019, Zhou2020a, RodriguezVillarreal2021} as well as numerically.\cite{Fedosov2010a, Yin2012, Katanov2015a, Balogh2019, Gracka2022} In general, the CFL in microchannels depends on various factors including the suspension hematocrit, channel dimensions, flow rate, and biophysical RBC properties such as their deformability.\cite{Kim2009, Tripathi2015} Additionally, the organization or focusing behavior of RBCs in dilute suspensions can also arise from geometric features of the channel, such as confinement,~\cite{Tomaiuolo2012, Iss2019} bifurcations,~\cite{Sherwood2014a, Shen2016, Bacher2018} and constrictions.~\cite{Faivre2006, Bacher2017, Abay2020} Sudden changes in the channel cross-section, similar to stenosed vessels, dramatically affect the RBC distribution and the CFL. RBC flow through simple and more sophisticated constriction geometries has been studied comprehensively under steady flow conditions foremost to mimic and understand pathological flow conditions in the microcirculation and to develop microfluidic plasma separation devices.\cite{Faivre2006, Fujiwara2009, KersaudyKerhoas2010a, RodriguezVillarreal2010a, Yaginuma2013, Pinho2013, Marchalot2014a, Rodrigues2016, Abay2020, RodriguezVillarreal2021, Gracka2022} In microfluidic constriction-expansion geometries, the constriction enhances the thickness of the CFL post-constriction. Here, the length and width of the constriction predominantly influence this CFL enhancement.\cite{Faivre2006, Yaginuma2013, Rodrigues2016, Gracka2022} Faivre~\etal~\cite{Faivre2006} showed that the growth of the downstream CFL under steady flow conditions further depends on the applied flow rate, suspending fluid viscosity, RBC concentration, and RBC deformability. Post-constriction, microfluidic expansions can generate large cell-free areas (CFA) with recirculation zones and vortices. This effect can be used for trapping or separating cells based on their rigidity and enables continuous plasma extraction from whole human blood in microfluidic devices.\cite{Sollier2010, Yang2012, Abay2020} Recently, Sanchez~\etal~\cite{Sanchez2022} investigated the interactions of RBCs with the expansion flow field of an in-vitro microfluidic model of a venous valve under steady flow conditions and finite inertia. Examining the local distribution of dilute RBC suspension passing the constriction, they showed that the local hematocrit profiles post-constriction are spatially heterogeneous and different from the feed hematocrit. 

So far, microfluidic studies have primarily focused on the flow behavior of RBC suspensions under steady flow conditions, since generating precise time-dependent flow conditions in microfluidic devices remains challenging.\cite{Recktenwald2021b} However, time-dependent flows play a crucial role in biomimicry in physiological studies, foremost to understand phenomena in the cardiovascular system, giving rise to unique nonlinear hydrodynamic instabilities and turbulence in unsteady flows.\cite{Xu2020} Additionally, time-dependent driving of the flow has attracted increasing attention since it enhances a broad range of operations and microfluidic applications, such as mixing, droplet generation, filtration of specific cells from whole blood, and biomechanical assessment of RBC properties.\cite{Ward2015, Yoon2016, Zhu2017, Dincau2020, Kang2023}

Nevertheless, detailed experimental studies on the time-dependent nature of RBC and CFL dynamics remain scarce. Common approaches to determining the CFL in experimental studies typically construct an RBC core flow in conjunction with the position of the channel walls.\cite{Balogh2019} From a technical perspective, the CFL is often determined by combining multiple hundred individual images obtained by high-speed video microscopy and generating the minimum intensity $z$-projection. With a sufficient number of image frames, the RBC core flow and the position of the channel walls can be detected, which allows calculating the CFL in a specific region of interest under steady flow conditions.\cite{Bugliarello1970, Kim2009} However, this method cannot be used so straightforwardly in the case of time-dependent flow if the flow dynamically changes during the image acquisition, especially for low-concentrated RBC suspensions. 

In this study, we, therefore, developed a technique that allows us to determine the cell-free zones during the periodic time-dependent flow of particle suspensions. We employ this method to examine the temporal CFL and CFA dynamics of RBCs in a constricted microfluidic device, covering a broad concentration range between \mbox{$\unit [1]{\%Ht}$} to \mbox{$\unit [20]{\%Ht}$}. We initially characterize the evolution of the CFL at multiple positions along the channel flow direction to quantify the effect of the applied pressure drop and hematocrit on the CFL and CFA pre- and post-constriction. Additionally, we examine the dynamics of the CFL and CFA at the constriction-expansion-region under a sinusoidal flow modulation for various pressure offsets and amplitudes. We further perform complementary 3D numerical simulations of a \mbox{$\unit [1]{\%Ht}$} RBC suspension both under steady and time-dependent flow conditions that allow us to resolve the time-dependent flow field at the constriction. Besides detailed descriptions of the effect of the hematocrit and the parameters of the applied pressure modulation of the CFL and CFA dynamics, our work reveals a phase inversion of the temporal CFA pre-constriction and provides a framework for studying time-dependent RBC and CFL phenomena even at low and high concentrations in complex microfluidic applications.

\section{Materials and methods}
\subsection{Experimental}
\subsubsection{Preparation of red blood cell suspensions}
Blood is taken with informed consent from healthy voluntary donors. It is centrifuged at \mbox{$\unit[1500]{g}$} for five minutes to separate RBCs and plasma. Sedimented RBCs are washed three times with phosphate-buffered saline solution (Gibco PBS, Fisher Scientific, Schwerte, Germany) before preparing RBC suspensions at hematocrit concentrations of \mbox{$\unit [1]{\%Ht}$}, \mbox{$\unit [5]{\%Ht}$}, \mbox{$\unit [10]{\%Ht}$}, and \mbox{$\unit [20]{\%Ht}$} in a PBS solution that contains \mbox{$\unit [1]{g\,L^{-1}}$} bovine serum albumin (BSA, Sigma-Aldrich, Taufkirchen, Germany). 

Blood withdrawal, sample preparation, and experiments were performed according to the guidelines of the Declaration of Helsinki and approved by the ethics committee of the `Aerztekammer des Saarlandes' (approval number 51/18).

\subsubsection{Microfluidic setup}
\begin{figure}[ht]
\centering
  \includegraphics[width=8.3cm]{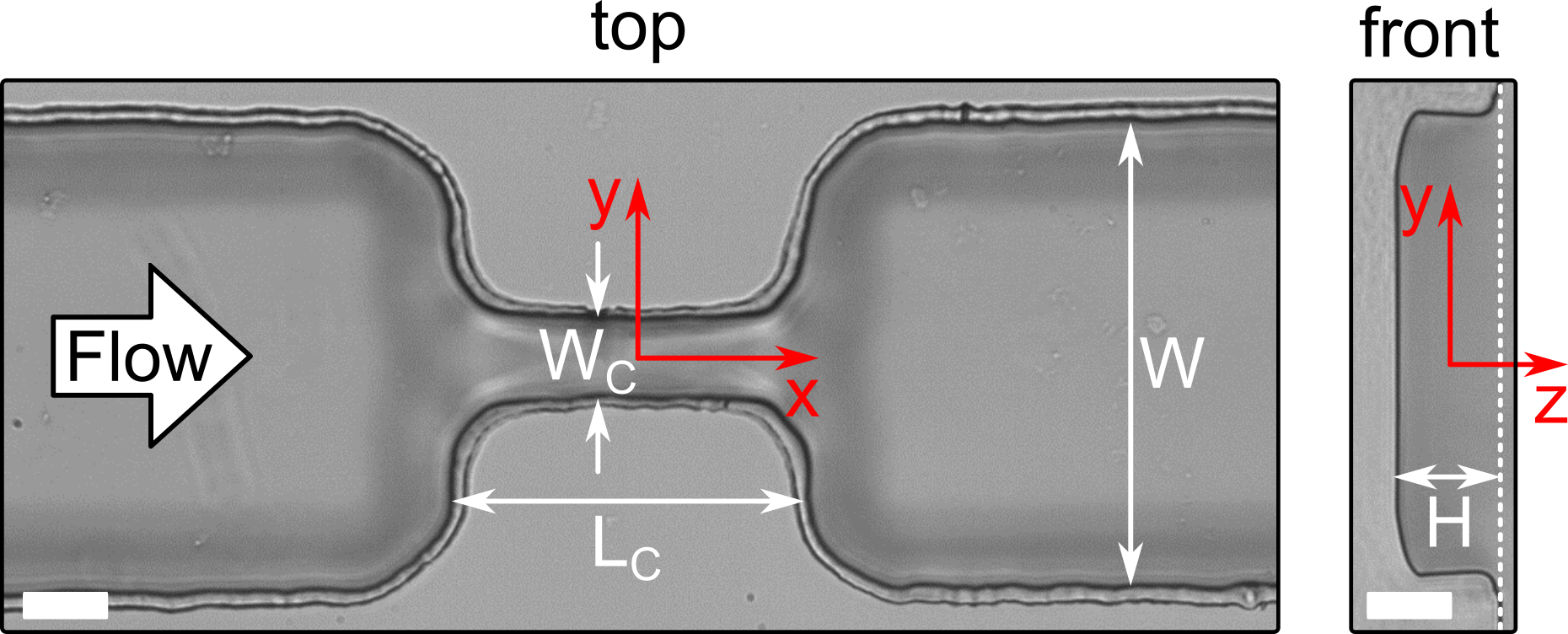}
  \caption{Top view (left) and front view (right) of the microfluidic constriction. The channel has a rectangular cross-section with width \mbox{$W=\unit[110]{\um}$} and height \mbox{$H=\unit[25]{\um}$}. The channel is constricted to a width of \mbox{$W_{\text{c}}=\unit[19]{\um}$} over a length of \mbox{$L_{\text{c}}=\unit[86]{\um}$}. Flow is from left to right. The dashed white line in the front view panel indicates the position of the glass slide that is bonded onto the PDMS chip. Scale bars represent \mbox{$\unit[20]{\um}$}. }
  \label{FIG_Setup}
\end{figure}

The RBC suspensions are pumped through a microfluidic channel with a high-precision pressure device (OB1-MK3, Elveflow, Paris, France). The microfluidic device is fabricated using polydimethylsiloxane (PDMS, RTV 615A/B, Momentive Performance Materials, Waterford, NY) through standard soft lithography.\cite{Friend2010} The straight microfluidic channel has a width of \mbox{$W=\unit[110]{\um}$} in $y$-direction, a height of \mbox{$H=\unit[25]{\um}$} in $z$-direction (aspect ratio \mbox{$\text{AR}=W/H=4.4$}), and a total length of \mbox{$L=\unit[20]{mm}$} in $x$-direction. The channel contains a constriction with a width of \mbox{$W_{\text{c}}=\unit[19]{\um}$} and a length of \mbox{$L_{\text{c}}=\unit[86]{\um}$} in the middle of the chip, as shown in Fig.~\ref{FIG_Setup}. The inlet and outlet are connected with rigid medical-grade polyethylene tubing (\mbox{$\unit[0.86]{mm}$} inner diameter, Scientific Commodities, Lake Havasu City, AZ) to the sample and waste containers, respectively. The microfluidic chip is mounted on an inverted microscope (Eclipse TE2000-S, Nikon, Melville, NY), equipped with LED illumination, a high-speed camera (Fastec HiSpec 2G, FASTEC Imaging, San Diego, CA), and a \mbox{$20\times$} air objective (Plan Fluor, Nikon, Melville, NY) with a numerical aperture \mbox{$\text{NA}=0.45$}.

In this study, we apply steady pressure drops as well as sinusoidal pressure modulations \mbox{$p(t) = p_0+p_\text{A} \sin(\omega t)$}, with pressure offset \mbox{$p_{0}$}, pressure amplitude \mbox{$p_\text{A}$}, and angular frequency \mbox{$\omega= 2\pi f$}. The frequency of the modulation is kept constant at \mbox{$f=\unit[1]{Hz}$}, while pressure offsets of \mbox{$p_{0}=\unit[250]{mbar}$}, \mbox{$\unit[375]{mbar}$}, and \mbox{$\unit[500]{mbar}$} in combination with three relative amplitudes \mbox{$p_\text{A}/p_{0}=\unit [20]{\%}$}, \mbox{$\unit [50]{\%}$}, and \mbox{$\unit [80]{\%}$} are used. The pressure modulation is applied for 40-60 periods and a global trigger signal is used to synchronize the pressure device and the high-speed camera.

\subsection{Data analysis}
\subsubsection{Determination of the CFL}
\label{sec:CFLdetermination}

\begin{figure*}[ht]
\centering
  \includegraphics[width=\textwidth]{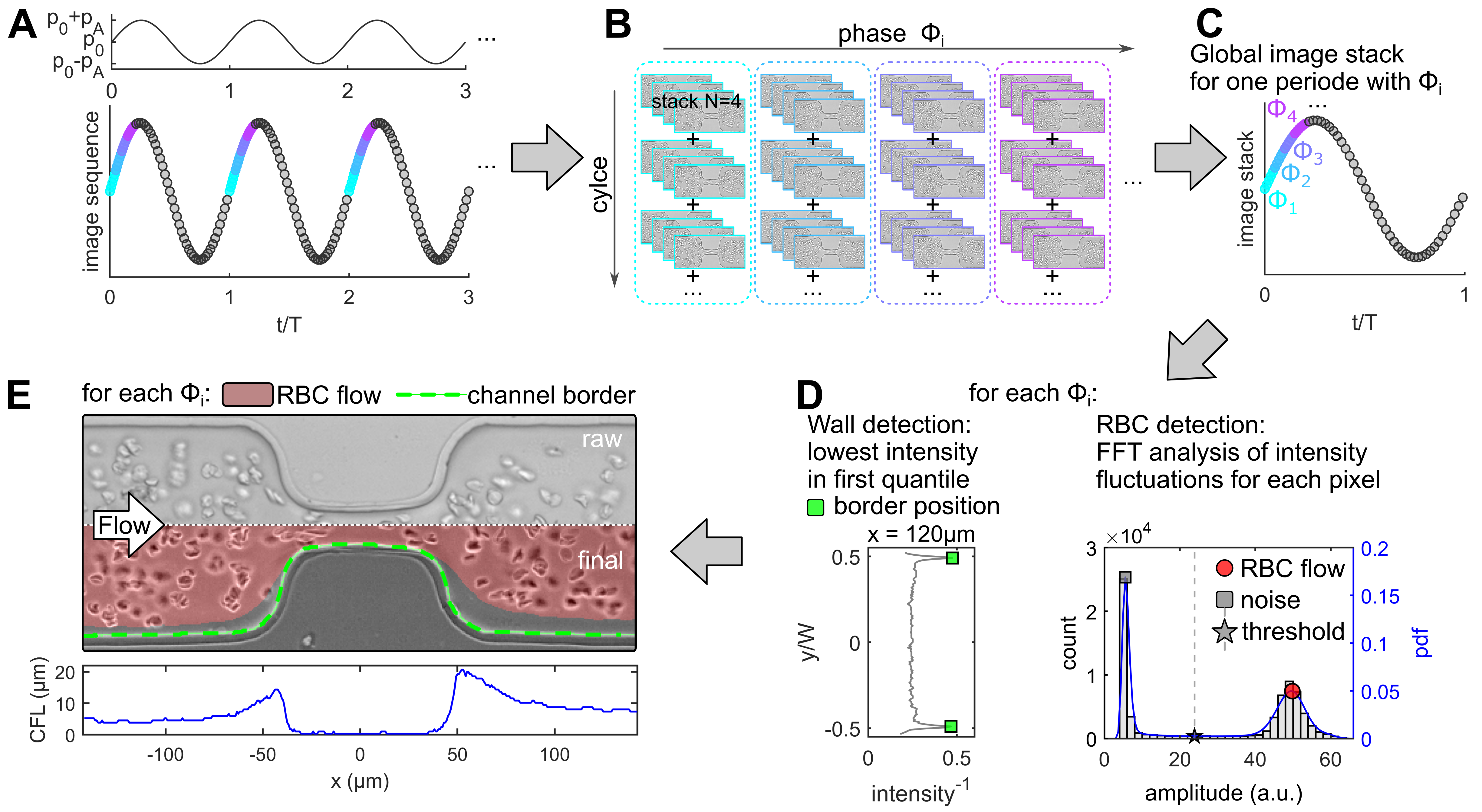}
  \caption{Detection method of the CFL in time-dependent RBC flow. (A) Sinusoidal pressure modulation (top) and schematic corresponding image sequence (bottom) for three periods. Colored symbols schematically indicate \mbox{$N=4$} images that are stacked and correspond to the same phase \mbox{$\phi_i$} within different periods. Note that the real-time over which \mbox{$N=4$} images are recorded is \mbox{$t=\unit[10]{ms}$}. (B) For each phase \mbox{$\phi_i$}, \mbox{$i=1 \dots 100$}, the individual image stacks over the 60 modulation cycles are stacked. (C) Resulting global image stack that covers one full period of the pressure modulation. For each phase \mbox{$\phi_i$} in this image stack statistical and Fourier methods are employed. (D) Representative histogram and pdf of the amplitudes of the intensity fluctuations in the FFT-integral image (right), yielding the threshold value to binarize the FFT-integral image and to detect the region of RBC flow. Representative plot of the inverted intensity as a function of the normalized channel width at \mbox{$x=\unit[120]{\um}$} (left). Here, the maximum positions correspond to the channel borders at that $x$-position, as indicated by the green symbols. (E) Raw image (top) of a \mbox{$\unit [5]{\%Ht}$} RBC suspension with superimposed RBC flow and border detection (bottom), representatively shown for one phase \mbox{$\phi_i$} during a pressure modulation with \mbox{$p_{0}=\unit[250]{mbar}$} and \mbox{$p_\text{A}=\unit[200]{mbar}$}. The lower panel in (E) shows the CFL between the RBC core flow and the border as a function of the $x$-position.}
  \label{FIG_Method}
\end{figure*}

In this work, we examine the CFL that is formed between the RBC core flow and the wall under steady and time-dependent flow conditions. For a periodic flow modulation, such as the used sinusoidal pressure signal, we have to account for the CFL changes over time. Therefore, we introduce a method that allows us to derive the time-dependent CFL for any arbitrary periodic waveform. To determine the CFL as a function of the corresponding phase, an image sequence is acquired over 40-60 periods during the pressure modulation, see Fig.~\ref{FIG_Method}A. A short exposure time of \mbox{$\unit[2]{ms}$} is used to avoid motion blur. The images are recorded at a frame rate of \mbox{$\unit[400]{fps}$}. In each cycle, \mbox{$N=4$} images are stacked at the same phase \mbox{$\phi_i$}. This results in the subdivision of the period into 100 steps. Therefore, the pressure and velocity changes during one image acquisition are negligible. This procedure is repeated for the whole period of the pressure modulation, as representatively shown in Fig.~\ref{FIG_Method}B. The image stacks belonging to the same phase \mbox{$\phi_i$} from the entire image sequence are combined into separate image stacks that cover one full period of the pressure modulation, see Fig.~\ref{FIG_Method}C. Those stacks are analyzed using statistical and Fourier methods, as schematically shown for one phase in Fig.~\ref{FIG_Method}D. In contrast to the calculation of the standard deviation of each pixel intensity, our FFT approach avoids false detection by slow illumination intensity fluctuations and is more sensitive to rare strong intensity changes, either if a cell passed the position of the corresponding pixel or a gap appears when continuous cells pass this pixel. This occurs at very small or very large cell concentrations, respectively. The intensity fluctuates strongly at pixel coordinates when cells pass this position over the time dimension of the image stack. However, the intensity remains almost constant outside these regions. We calculate the temporal Fourier transform (FFT) for each pixel intensity to quantify those local fluctuations over time. The integral over the upper half, the high-frequency part is calculated to obtain an FFT-integral matrix with the same dimensions as the original image. An automated, robust method is used to threshold this integral image. The calculated probability density distribution (pdf) by kernel density estimation of all amplitudes in the integral image is a bimodal distribution, as shown in the right panel in Fig.~\ref{FIG_Method}D. One peak corresponds to small fluctuation amplitudes due to camera noise at almost constant pixel intensities over time, while the other peak corresponds to large amplitudes due to the passing of cells. The local minima in the pdf between those peaks are used as a threshold to binarize the FFT-integral image and the corresponding pixel positions represent the location of RBC flow, see red shaded areas in Fig.\ref{FIG_Method}E. To detect the borders of the channel at a certain phase of the pulsation, the first quantile is calculated from the associated image stack to avoid outliers. The lowest intensity in $y$-direction for each position along the $x$-direction results in the wall position, schematically shown at an $x$-position of \mbox{$x=\unit[120]{\um}$} in the left panel of Fig.~\ref{FIG_Method}D. We find that the walls only pulse less than one pixel during a pulsation period with a small amplitude at a pixel position on the wall in phase with the applied pressure. The maximum wall deformation that we observe for the highest applied pressure drop combination, \textit{i.e.}, \mbox{$p_{0}=\unit[500]{mbar}$} and \mbox{$p_\text{A}=\unit[400]{mbar}$}, is \mbox{$\unit[0.37]{\um}$}, as shown in Fig.~S1 in the supplementary material. Hence, we establish this value as a conservative lower detection limit for our experimental investigations regarding the CFL under time-dependent flow conditions. Figure~\ref{FIG_Method}E representatively shows the result of the RBC and border detection for one phase during the period. The red-shaded area corresponds to the regions of RBC flow, while the green-dashed line indicates the channel borders. The vertical distance between the region with the detected cell and the position of the walls yields the \mbox{$\text{CFL}(x)$} for each $x$-position, as shown in the lower panel of Fig.~\ref{FIG_Method}E. We apply this method to experimental data to extract the time-dependent \mbox{$\text{CFL}(x)$} along the flow direction.

\subsubsection{Flow velocimetry}
We employ particle image velocimetry (PIV) at \mbox{$x=\unit[-5]{mm}$} pre-constriction to determine the suspension velocity at a given pressure drop. RBCs are used as flow tracers, without the addition of further particles.\cite{Poelma2012, Passos2019, Abay2020} The measurement depth $\delta z_{\text{m}}$ over which cells are detected and contribute to the determination of the velocity field is $\delta z_{\text{m}}\approx44~\mu$m for the $20\times$ lens.\cite{Meinhart2000} Hence, the RBC velocity is averaged over the depth of the channel ($\delta z_{\text{m}}/H\approx1.76$). An open-source PIV software\cite{Thielicke2014a} is used to calculate the velocity profile across the channel width. Representative velocity profiles and the maximum velocity \mbox{$v_{\text{max}}$} for the RBC suspensions at different concentrations and pressure drops are shown in Fig.~S2 in the supplementary material.

Based on the velocity at $y=0$ \mbox{$v_{\text{max}}$} at a given pressure drop, we calculate the Reynolds number \mbox{$\text{Re}$}, which relates the inertial to viscous forces in the system,  as \mbox{$\text{Re} = {v_{\text{max}} D_{\mathrm{h}}\rho}/{\eta}$}, with the fluid density \mbox{$\rho=\unit[1]{g\,cm^{-3}}$}, the fluid's dynamic viscosity \mbox{$\eta$}, and the hydraulic diameter of the rectangular microfluidic channel \mbox{$D_{\mathrm{h}}=2W H/(W+H)$}. We perform steady shear measurements according to Horner~\etal~\cite{Horner2019} to assess the RBC suspension viscosity. Therefore, a stress-controlled rheometer (MCR702, Anton Paar, Graz, Austria) equipped with a coaxial cylinder geometry (CC20, Anton Paar, Graz, Austria, inner diameter \mbox{$r_{\text{i}}=\unit[10]{mm}$}, radius ratio \mbox{$\delta=r_{\text{a}}/r_{\text{i}}=1.1$}) is used at a constant temperature of \mbox{$T=\unit[20]{^\circ C}$}. The viscosity as a function of the shear rate for the different suspensions is shown in Fig.~S3 in the supplementary material. At high shear rates above \mbox{$\dot\gamma=\unit[10]{s^{-1}}$}, the viscosity is constant and slightly increases with RBC concentration, as shown in the inset in Fig.~S3. For the microfluidic chip, we estimate the nominal wall shear rate in the straight channel as \mbox{$\dot \gamma \approx 6v_{\text{max}}/H \gg \unit[1000]{s^{-1}}$}. Hence, we use the viscosity values listed in Table~S1 in the supplementary material to calculate \mbox{$\text{Re}$} for each RBC suspension and applied pressure drop.

In pulsatile flows, the dimensionless Womersley number\cite{Womersley1955} \mbox{$\text{Wo}$} relates transient inertia effects to viscous forces and is defined as \mbox{$\text{Wo} = {D_{\mathrm{h}}}/{2}\,{\sqrt{{\omega \,\rho}/{\eta}}}$}. For \mbox{$\text{Wo}\ll 1$}, viscous effects dominate the flow and the pulsation frequency is small enough for the steady velocity profile to develop. However, when \mbox{$\text{Wo}> 1$}, the transient inertia forces dominate the flow dynamics, which can result in strong deviations of the mean velocity from the Poiseuille flow profile. This effect can lead to flow reversal near the channel walls in pulsatile flows.\cite{Blythman2017} In the experiments performed here, we find \mbox{$\text{Wo}\approx0.05$}, hence, the influence of transient inertia forces on the flow can be neglected.

\subsection{Numerical simulations}
We perform numerical simulations of the RBCs at \mbox{$\unit [1]{\%Ht}$} in a constricted channel under steady and time-dependent flow at similar \mbox{$\text{Re}$} as in the experiments. To avoid the instabilities of the triangulated cell membranes that can emerge at high shear rates at the constriction, we increase the dimensions of the channel geometry by a factor of two in the numerical simulations compared to the microfluidic channel. Hence, the channel in the simulations has a constant height of \mbox{$H=\unit [50]{\um}$} and width of \mbox{$W=\unit [211]{\um}$} before and after the constriction. The total length of  the simulated channel is \mbox{$L=\unit [832]{\um}$}. The constriction is \mbox{$L_{\text{c}}=\unit [142]{\um}$} long and has a width of \mbox{$W_{\text{c}}=\unit[38]{\um}$}, as shown in Fig.~S4 in the supplementary material. 

To mimic the inflow from the fluid reservoir in the experiments, cells are inserted with randomized input positions at the channel entrance. The feed-in mechanism is constructed such that the hematocrit at the channel entrance is \mbox{$\unit [1]{\%Ht}$}, similar to the experimental situation. The cells are removed from the simulations when they  reach the channel outlet. 

The RBCs are modeled as two-dimensional elastic membranes, which show resistance to shear and area dilation implemented with Skalak's law,\cite{Skalak1973, Barthes2002} and also resistance to bending incorporated with the Helfrich law\cite{Canham1970, Helfrich1973, Guckenberger2016, Guckenberger2017}. The shear modulus is set to \mbox{$\kappa_\mathrm{S}= \unit[25\times10^{-6}]{N/m}$}, which is larger by a factor 3 to 5 compared to the values found in the literature.\cite{Hochmuth1987, Mills2004, Yoon2008} While this modification is expected to have negligible influence on the collective cell behavior, it increases the stability of the cell under high shear stresses, especially at the constriction entrance, which is necessary to reach the required velocities and \mbox{$\text{Re}$}. The empirical modulus for the area dilation is set to  \mbox{$\kappa_\mathrm{A}=40\times\kappa_\mathrm{S}$} and the bending modulus is \mbox{$\kappa_\mathrm{B}=\unit[2\times10^{-19}]{N\,m}$},\cite{Evans1983, Freund2014} with a flat bending reference shape. The long radius of the simulated RBCs in their characteristic biconcave shape at rest is \mbox{$R=\unit[3.91]{\um}$}. Their surface area is \mbox{$A=\unit[133.5]{\um^2}$} and their volume \mbox{$V=\unit[93.5]{\um^3}$}.\cite{Evans1972, Diez2010, Skalak1989}. The RBC surface is discretized with 162 flat triangles per cell in our numerical simulations. The fluid inside the channel is simulated as a Newtonian fluid with a dynamic viscosity of \mbox{$\eta=\unit[1.2]{mPa\,s}$},\cite{Chien1966} and a density of \mbox{$\rho=\unit[1000]{kg/m^3}$}.\cite{Bronzino2006} RBCs are filled with a Newtonian fluid with the same viscosity as the surrounding fluid.

A three-dimensional lattice Boltzmann method (LBM) is used to simulate the fluid flow through the channel.\cite{Kruger2017, Dunweg2009, Aidun2010} We use the implementation of the LBM in the software package ESPResSo.\cite{Limbach2006a, Arnold2013, Weik2019} The cell membranes are coupled to the flow with the immersed boundary method (IBM).\cite{Peskin2002, Mittal2005, Kruger2011}.

Simulations under steady flow conditions are performed at flow velocities of \mbox{$\unit [80]{mm/s}$}, \mbox{$\unit [120]{mm/s}$}, and \mbox{$\unit [160]{mm/s}$}, which results in \mbox{$\text{Re}\approx5.3$}, 8, and 11, respectively. In the numerical simulations, \mbox{$\text{Re}$} is calculated with the same equation as for the experiments and is on the same order of magnitude as the experiments for \mbox{$\unit [1]{\%Ht}$} (see Table~S1). We further simulate flow through the constriction for a sinusoidal flow oscillation with a constant frequency of \mbox{$f=\unit [20]{Hz}$} (\mbox{$\text{Wo}\approx0.4$}). We did not simulate smaller frequencies near the experimental \mbox{$f=\unit [1]{Hz}$} because this would lead to a large increase in computation time. Representative velocity profiles pre-constriction are shown in Fig.~S5 in the supplementary material, which doesn't indicate any influence of transient inertia forces on the flow.

To compute the CFL in the numerical simulations under steady flow conditions, we use the same image stacking method as for the experiments (see Fig.~\ref{FIG_Method}) based on the graphical representation of the simulation data. Additionally, we used a second method based on the numerical data of the discretized RBC surface. For this method, we divide the channel into bins with a width of approximately \mbox{$\unit[10]{\um}$} in $x$-direction. In each bin, the RBC surface node that is nearest to the channel wall is used to calculate the distance to the wall. This process is repeated for every frame in the numerical simulation and the minimum distance for each bin in the $x$-direction over the sequence is used as \mbox{$\text{CFL}(x)$}. For further reference, we refer to this numerical method as the cell contour method. For the time-dependent flow simulations, the limited number of periods does not allow a straightforward application of the image stacking method to determine the time-dependent CFL. Therefore, each period of the flow oscillation in the numerical simulation is divided into 50 time bins. At each time bin that corresponds to the same phase for different periods, we determine the minimum distance to the wall for each $x$-position, hence \mbox{$\text{CFL}(x)$} for each time bin.

\section{Results and discussion}

\subsection{Investigations under steady flow conditions}
\subsubsection{RBC distribution along the flow direction}

\begin{figure*}[ht]
\centering
  \includegraphics[width=\textwidth]{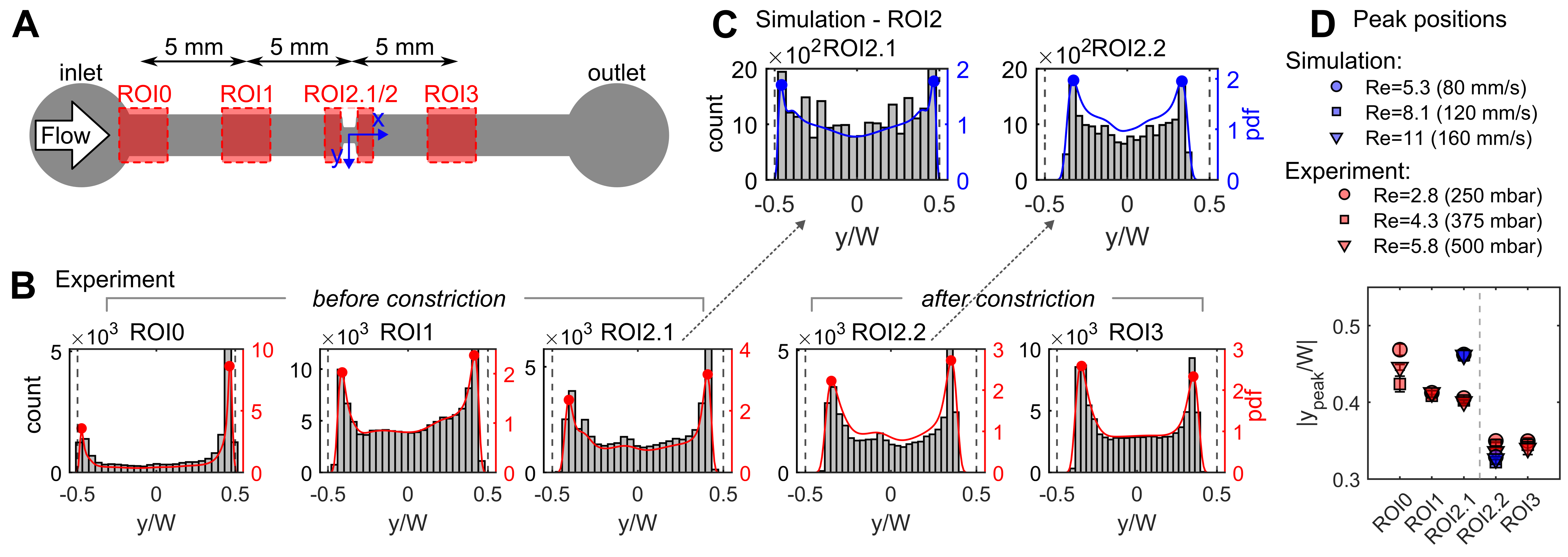}
  \caption{RBC distributions of a \mbox{$\unit[1]{\%Ht}$} RBC suspension under constant flow conditions. (A) Schematic representation of the ROIs in the experiments. (B) RBC distribution across the normalized channel width in the ROIs for the experiment at \mbox{$\unit[250]{mbar}$} (\mbox{$\text{Re}= 2.8$}). Red lines show the probability density functions (pdfs) and vertical dashed lines indicate the channel walls. (C) RBC distribution across the normalized channel width at the constriction in ROI2 for the simulation at \mbox{$v=\unit[80]{mm/s}$} (\mbox{$\text{Re}= 5.3$}). (D) Absolute normalized peak positions of the RBC distributions in the different ROIs for various \mbox{$\text{Re}$}. The dashed vertical line separates the pre-constriction positions from the post-constriction positions.}
  \label{FIG_steadyDist}
\end{figure*}

First, we examine the development of the RBC distribution along the channel flow direction to understand the spatiotemporal RBC organization under steady flow conditions. Therefore, we experimentally probe the RBC flow of a  \mbox{$\unit[1]{\%Ht}$} RBC suspension in five specific regions of interest (ROIs), schematically shown in Fig.~\ref{FIG_steadyDist}A. Figure~\ref{FIG_steadyDist}B shows the RBC distributions across the channel width \mbox{$W$} and corresponding probability density functions (pdfs) for the experiments at a constant pressure of \mbox{$\unit[250]{mbar}$}, corresponding to a maximum velocity of \mbox{$v_{\text{max}}=\unit[80]{mm/s}$} and \mbox{$\text{Re}= 2.7$}. Directly after the inlet reservoir of the microfluidic chip at ROI0, we observe an RBC distribution, characterized by the emergence of two peaks close to the channel walls while fewer RBCs flow in the channel center. In our experiments, the peculiar RBC distribution is already formed within the first few \mbox{$\unit[100]{\um}$} in the straight channel, as shown in Fig.~S6 in the supplementary material. Hence, it forms as a consequence of the inlet flow from the larger circular inlet reservoir, which has a diameter of \mbox{$\unit[2]{mm}$} and leads into the straight channel with a narrower width of \mbox{$W=\unit[110]{\um}$}. With increasing distance from the fluid inlet, the qualitative shape of the RBC distribution persists both pre-constriction (ROI1 and ROI2.1) as well as post-constriction (ROI2.2 and ROI3). At the same time, the formation of the pronounced cell-free region close to the walls post-constriction is visible in the histograms.

The RBC distributions before and after the constriction in the simulation (corresponding to ROI2.1 and ROI2.2) is shown in Fig.~\ref{FIG_steadyDist}C. In our simulations, we feed the cells at a distance of roughly \mbox{$\unit[410]{\um}$} before the constriction. Therefore, cells do not travel a distance of roughly \mbox{$\unit[10]{mm}$} before the constriction as in the experiment and therefore, RBCs flow closer to the walls in ROI2.1 in the simulations. Post-constriction, the hematocrit profile persists, and a large cell-free zone is generated in the vicinity of the channel wall in agreement with the experiment, as shown in ROI2.2 in Fig.~\ref{FIG_steadyDist}C.

The absolute values of the normalized peak positions in the RBC distributions of Fig.~\ref{FIG_steadyDist}B and C are summarized in Fig.~\ref{FIG_steadyDist}D. RBCs accumulate at \mbox{$|y/W|\approx0.44$} entering the channel in ROI0. Further downstream, the peak positions move slightly inwards in the experiments to \mbox{$|y/W|\approx0.41$} and \mbox{$0.4$} in ROI1 and ROI2.1, respectively. Post-constriction, the peak positions keep moving inward towards the channel center at \mbox{$|y/W|\approx0.34$}, similar to previous experimental studies.\cite{Faivre2006} These observed peak positions of RBC focusing in the different ROIs do not depend on the applied pressure drop within the investigated range. For the numerical simulations, we find similar peak positions after the constriction while the peak position before the constriction is closer to the wall because the ROI2.1 in the simulations is directly behind the feed-in region.

Similar characteristic cell distributions, manifested by an off-center two-peak profile (OCTP), have also been found for RBC suspensions in straight rectangular channels, however, at negligible inertia. Zhou~\etal~\cite{Zhou2020a} recently investigated the spatiotemporal dynamics of dilute RBC suspensions at \mbox{$\text{Re}< 2\times 10^{-4}$} using numerical simulations as well as microfluidic experiments. In a similar straight rectangular channel (\mbox{$\text{AR}=3.2$}), they observed the formation of OCTPs with two peaks around \mbox{$|y/W|\approx0.35-0.4$}, saturating after roughly \mbox{$\unit[1-2]{mm}$} downstream. The authors found that the hematocrit profiles were predominantly determined by the decay of hydrodynamic lift within the suspension. Furthermore, they showed that the initial inflow configuration in the channel codetermined the observed RBC ordering phenomenon.\cite{Zhou2020a} 

In low-Reynolds-number microfluidic shear flows, the cross-streamline migration of RBCs is induced by mainly induced by three mechanisms: 1) interactions with the boundary walls that generate a repulsion lift force on the RBCs toward the channel center, 2) shear-gradient-induced motion induced by flow profile curvature, which also pushes the RBCs in the center direction, and 3) hydrodynamic interactions between the cells, known as shear-induced dispersion, whose net effect is to produce motion down the concentration gradient, toward the channel wall.\cite{Secomb2017} However, in the results shown in Fig.~\ref{FIG_steadyDist}, the two-peaked RBC concentration profiles at \mbox{$\text{Re}\approx2$} already emerge directly after the inlet in ROI0. Therefore, the observed hematocrit profiles rather manifest as a consequence of the initial cell distribution in the larger reservoirs, which was shown to impact the downstream RBC distribution.\cite{Zhou2020a}

\subsubsection{RBC flow and CFL at the constriction}

\begin{figure}[ht]
\centering
  \includegraphics[width=8.1cm]{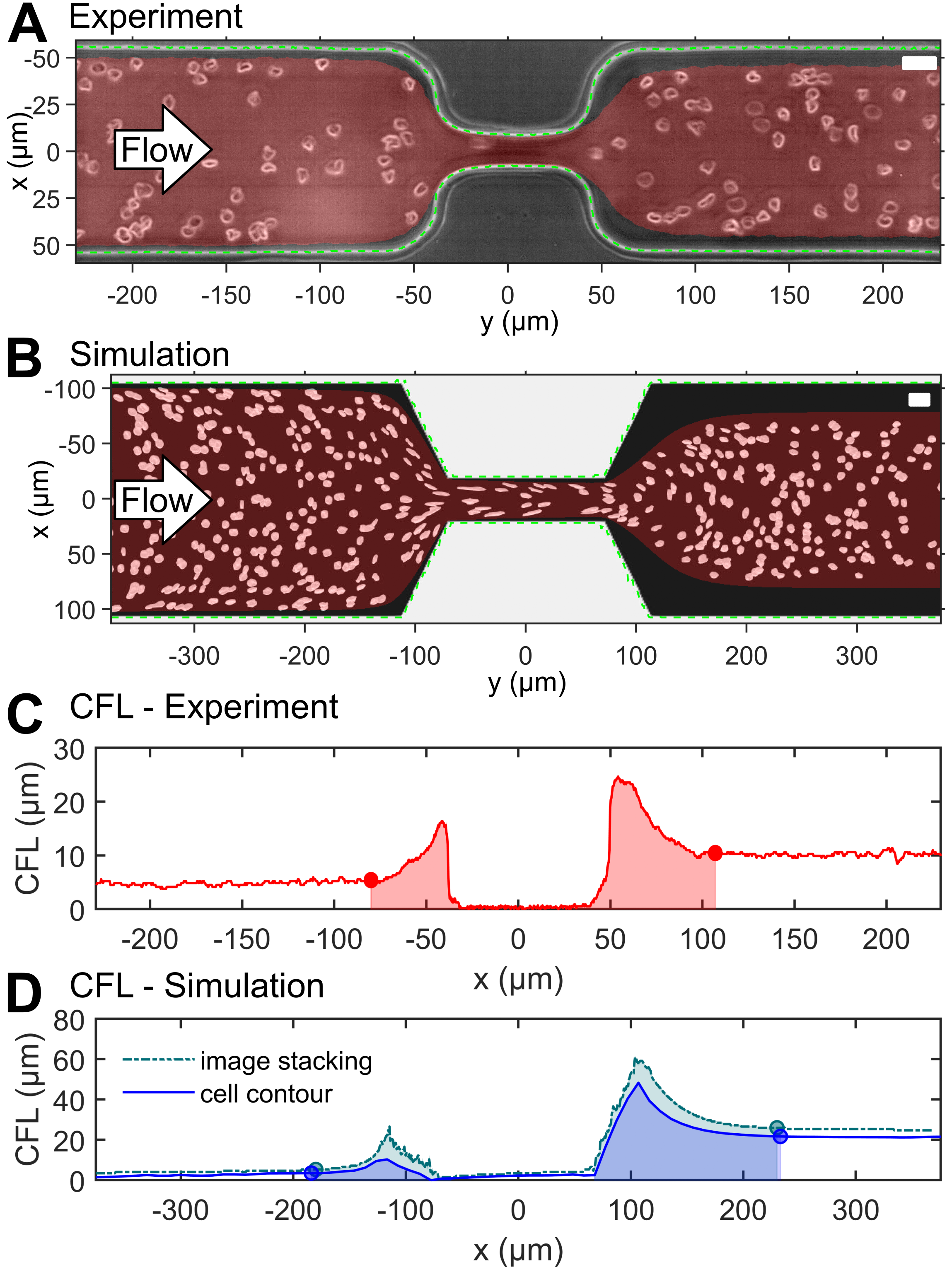}
  \caption{RBC flow of a \mbox{$\unit[1]{\%Ht}$} RBC suspension under constant flow conditions through the constriction. (A) and (B) representative snapshots of the RBC flow at the constriction for an experiment at \mbox{$\unit[250]{mbar}$} and a simulation with \mbox{$v=\unit[80]{mm/s}$}, respectively. The RBC core flow is indicated by the red area and the channel borders are highlighted by green dashed lines. Scale bars represent \mbox{$\unit[20]{\um}$}. CFL along the $x$-direction at the constriction for (C) the experiment and (D) the simulation shown in (A) and (B), respectively. The shaded areas in (C) and (D) indicate the CFA post and pre-constriction. Circular symbols in (C) and (D) correspond to the transition points in $x$-direction between the CFL and the CFA regions.}
  \label{FIG_steadyCFL}
\end{figure}

To understand the effect of a time-dependent flow modulation on the RBC flow behavior and the CFL dynamics at the constriction, we first examine how the sudden change of the channel cross-section affects the RBC organization under steady flow conditions. Figure~\ref{FIG_steadyCFL} shows representative snapshots for the flow of a \mbox{$\unit[1]{\%Ht}$} RBC suspension under constant flow at \mbox{$\unit[250]{mbar}$} (\mbox{$\text{Re}=2.8$}) for (A) an experiment and (B) a corresponding numerical simulation with \mbox{$v=\unit[80]{mm/s}$} (\mbox{$\text{Re}=5.3$}). 

The thickness of the CFL along the flow direction for the data of Fig.~\ref{FIG_steadyCFL}A are plotted in Fig.~\ref{FIG_steadyCFL}C. Before the constriction at \mbox{$x\leq\unit[-75]{\um}$} (corresponding to ROI2.1) we observe a constant CFL of roughly \mbox{$\unit[5.3]{\um}$} in the experiment. Between \mbox{$\unit[-75]{\um}\leq x<\unit[-40]{\um}$}, we observe an increase of the CFL before the throat of the contraction. Passing the narrow constriction, the RBC flow in close proximity to the channel wall. Thus, we do not detect a CFL in the passage region between \mbox{$\unit[-40]{\um}\leq x<\unit[40]{\um}$}. Subsequently, the strong jet flow region results in a sharp increase in the CFL between \mbox{$\unit[40]{\um}\leq x<\unit[100]{\um}$} after the channel expansion. For \mbox{$x\geq\unit[100]{\um}$}, the CFL plateaus at \mbox{$\unit[10.5]{\um}$}, corresponding to a doubling of the CFL thickness pre-constriction. 

In the numerical simulations, we observe a qualitatively similar behavior of the CFL, as shown in Fig.~\ref{FIG_steadyCFL}D. Here, we use both the image stacking method as well as the cell-contour method to extract \mbox{$\text{CFL}(x)$}. Before the constriction at \mbox{$x\leq\unit[-180]{\um}$}, the CFL is very small. However, it shows a slight increase along the flow direction as the RBCs are pushed away from the channel walls after the feed-in region. Before the constriction, \mbox{$\unit[-180]{\um}\leq x<\unit[-80]{\um}$}, the CFL increases before it approaches zero at the entrance of the constriction. In contrast to the experiments, we observe a slight increase in the CFL in the narrow constriction between \mbox{$\unit[-80]{\um}\leq x<\unit[80]{\um}$}, possibly due to the larger relative constriction size in the simulations with respect to the RBC size. Post constriction, the numerical simulations also show an increase in \mbox{$\text{CFL}(x)$} before reaching a stable plateau value. The shape of \mbox{$\text{CFL}(x)$} in the simulations is slightly different from the experiments as a result of the different corner and edge shapes of the constrictions. Further, the cell contour method yields a smaller \mbox{$\text{CFL}(x)$}. In this method, each RBC contributes to the \mbox{$\text{CFL}(x)$} detection. Hence, if a single RBC flows at a streamline close to the walls, it determines the minimum \mbox{$\text{CFL}(x)$} at that $x$-position. In contrast, if such a singular event occurs in the experiments, \textit{e.g.}, due to a single RBC with impaired deformability that flows closer to the channel wall at a given $x$-position, it has only negligible impact on the FFT analysis at this pixel position (see Fig.~\ref{FIG_Method}D). Hence, this position is not detected as part of the RBC core flow region, resulting in a larger \mbox{$\text{CFL}(x)$} with the image stacking method as compared to the cell contour method, also if the image stacking method is applied to the simulation data.

The shaded areas in Fig.~\ref{FIG_steadyCFL}C and D correspond to the CFAs pre- and post-constriction. The CFA post-constriction starts when the CFL before the constriction deviates by more than \mbox{$\unit[10]{\%}$} from its mean value and ends at the entrance of the constriction. Similarly, the CFA post-constriction starts at the constriction exit and ends when a stable value of the CFL post-constriction is reached. These transition points are highlighted with circular symbols in Fig.~\ref{FIG_steadyCFL}C and D. Note that the CFL in the numerical simulations, especially post-constriction, is larger by a factor of roughly two compared to the microfluidic experiments due to the larger size of the channel in the simulations.

\subsubsection{Effect of RBC concentration and pressure drop on the CFL and CFA}
\begin{figure}[ht]
\centering
  \includegraphics[width=8.1cm]{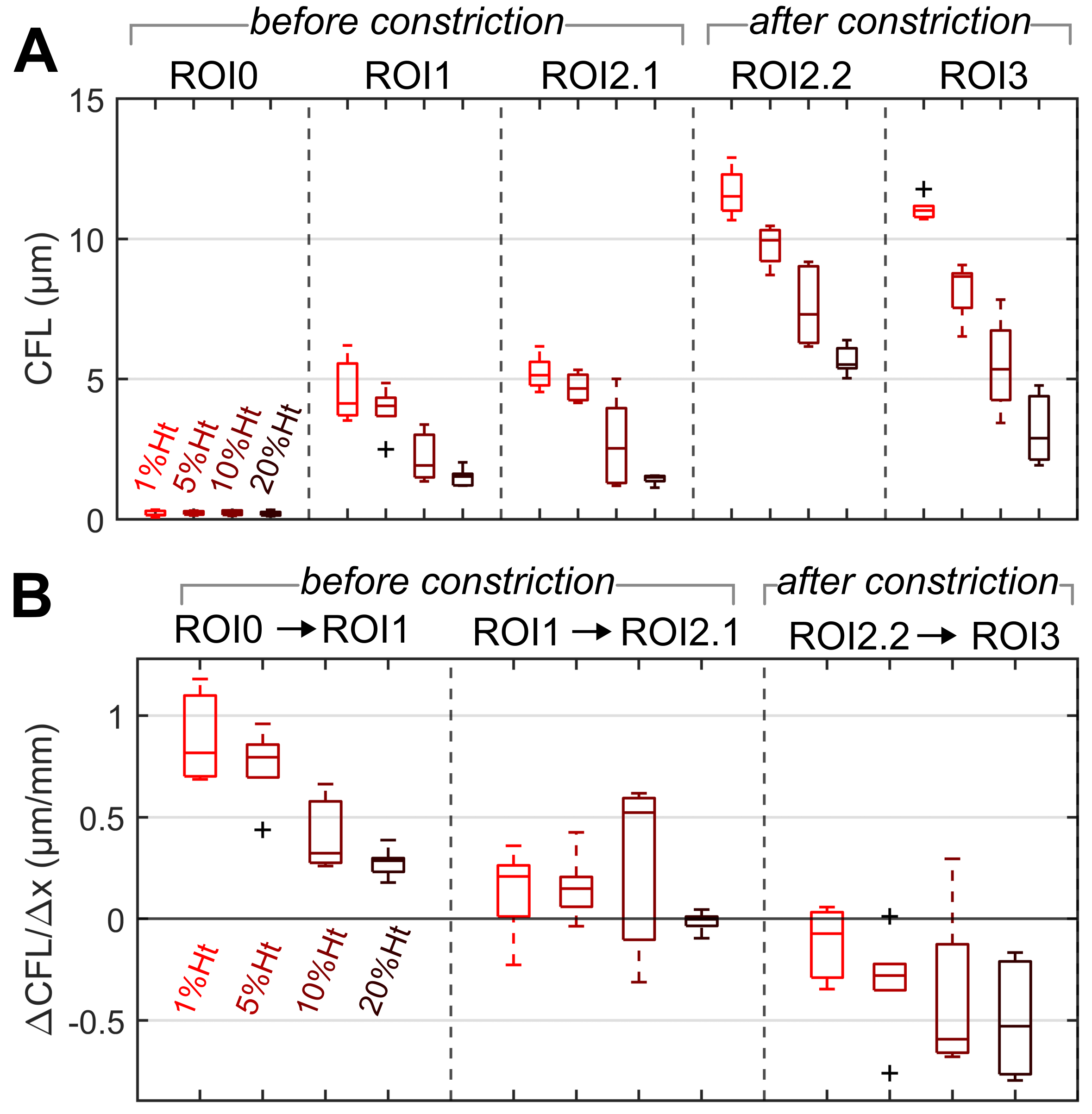}
  \caption{Effect of the RBC concentration on the CFL in the experiments at a constant pressure drop of \mbox{$p=\unit[500]{mbar}$}. (A) The thickness of the CFL for different RBC concentrations in the different ROIs (see Fig.~\ref{FIG_steadyDist}A) before and after the constriction. (B) Build-up rate of the CFL (\mbox{$\Delta \text{CFL} / \Delta x$}) between the different ROIs in the straight parts of the channel before and after the constriction.}
  \label{FIG_steadyCFLHt}
\end{figure}

In the microfluidic experiments, we further investigate the influence of the RBC concentration on the magnitude of the CFL and CFA under steady flow conditions. Figure~\ref{FIG_steadyCFLHt}A shows the CFL in the five different ROIs (see Fig.~\ref{FIG_steadyDist}A) for hematocrits of \mbox{$\unit[1]{\%Ht}$}, \mbox{$\unit[5]{\%Ht}$}, \mbox{$\unit[10]{\%Ht}$}, and \mbox{$\unit[20]{\%Ht}$} at a constant pressure drop of \mbox{$\unit[500]{mbar}$}. Directly after the inlet (ROI0), the CFL is zero for all investigated RBC concentrations (see Fig.~S6 for \mbox{$\unit[1]{\%Ht}$}). Subsequently, we observe a build-up of the CFL pre-constriction at ROI1  with an average value of \mbox{$\unit[4.5]{\um}$} for \mbox{$\unit[1]{\%Ht}$}, while we find a CFL of merely \mbox{$\unit[2]{\um}$} for the \mbox{$\unit[20]{\%Ht}$} suspension. Afterward, the thickness increases only minor within the next \mbox{$\unit[5]{mm}$} in ROI2.1 right before the constriction, apparently reaching a plateau value corresponding to an equilibrium CFL thickness in the experiments. Post-constriction, we observe a pronounced increase in the CFL for all RBC concentrations. For \mbox{$\unit[1]{\%Ht}$}, the CFL after the constriction increases to approximately \mbox{$\unit[11.5]{\um}$}, while the CFL for the \mbox{$\unit[20]{\%Ht}$} sample increases to roughly \mbox{$\unit[6]{\um}$}. The ratio between the CFL post/pre-constriction increases with the suspension hematocrit from a value of two to roughly four between \mbox{$\unit[1]{\%Ht}$} and \mbox{$\unit[20]{\%Ht}$}, as shown in Fig.~S7A in the supplementary material. Moreover, this ratio seems to be independent of the applied pressure drop within the investigated range. Further downstream of the constriction in ROI3, the CFL decreases again, as shown in Fig.~\ref{FIG_steadyCFLHt}A. Increasing the hematocrit leads to a decrease in the thickness of the CFL in each ROI, in accordance with previous investigations.\cite{Fedosov2010a, Tripathi2015, Bento2019} Furthermore, we do not observe a significant influence of the applied pressure drop within the investigated range on the thickness of the CFL in the different ROIs pre-constriction, similar to previous studies that observed only very little variation of the CFL with flow rate.\cite{Faivre2006, Tripathi2015} Merely for concentrations \mbox{$\leq \unit[5]{\%Ht}$}, we observe a slight increase of the CFL post-constriction with the pressure drop, as shown in Fig.~S8 in the supplementary material. Since the CFL post-constriction is larger than the equilibrium CFL, RBCs at lower velocities have more time to flow back toward the channel walls after the constriction.

To assess the rate at which the CFL develops in the straight parts of the microfluidic channel, we calculate the difference between the CFL in consecutive ROIs divided by their distance in $x$-direction \mbox{$\Delta \text{CFL} / \Delta x$}. Figure~\ref{FIG_steadyCFLHt}B shows this rate for the different RBC concentrations at a constant pressure drop of \mbox{$\unit[500]{mbar}$}. During the passage from the inlet (ROI0) to ROI1, all RBC suspensions show a build-up of the CFL. This build-up rate decreases with increasing hematocrit. In the second part of the straight channel pre-constriction, between ROI1 and ROI2.1, the CFL exhibits a much small increase, while some measurements actually show a saturation (\mbox{$\Delta \text{CFL} / \Delta x\approx0$}) or decrease of the CFL (\mbox{$\Delta \text{CFL} / \Delta x<0$}). In the straight channel section post-constriction, the thickness of the CFL decreases again, where this decrease is faster (\mbox{$\Delta \text{CFL} / \Delta x<0$}) the higher the RBC concentration. This is in striking contrast to the observations between ROI0 and ROI1, where the lowest hematocrit show the fastest development. The faster development of the CFL for high hematocrits in the straight channel post-constriction is due to the enhancement of RBC collisions, which push the strongly accumulated RBCs in the RBC core toward the wall and toward their equilibrium position again. In contrast, these collisions at higher RBC concentrations hinder the formation of a pronounced CFL pre-constriction in ROI1 and ROI2.1. Hence, we observe a large absolute CLF modification \mbox{$\lvert \Delta \text{CFL} / \Delta x\rvert$} for small hematocrit pre-constriction but for high hematocrit post-constriction. Moreover, the slow decrease of the CFL post-constriction for lower RBC concentrations is in qualitative agreement with previous studies, which showed that the geometry-induced enhancement of the CFL can last more than \mbox{$\unit[1]{cm}$} downstream of the constriction.\cite{Faivre2006, Bacher2018} Note, that the rate at which the CFL changes along the flow direction does not depend on applied pressure drop within the investigated range, as shown in Fig.~S9 in the supplementary material. However, we only use a factor of two between the smallest and largest applied pressure in this study. Using a larger pressure range would eventually result in a dependency of the CFL on \mbox{$p$}, as reported elsewhere.\cite{Abay2020} 

Our experimental observations pre-constriction are similar to equilibrium CFLs reported in other experimental studies in straight pipes and channels.\cite{Marchalot2014a}. However, Zhou~\etal~\cite{Zhou2020a} recently investigated the CFL growth under steady flow conditions in a comparable straight rectangular channel (\mbox{$\text{AR}=3.2$}) at \mbox{$\text{Re} \ll 1$} using numerical simulations and microfluidic experiments. At low RBC concentrations, they observed a build-up of the CFL following a power-law behavior in the simulations. The CFL increased up to \mbox{$\unit[6-9]{\um}$} over a length of \mbox{$28\times L/D_{\mathrm{h}}$}. However, the authors found a much slower CFL build-up in the experiments, which lasted until their maximum detection distance of \mbox{$46\times L/D_{\mathrm{h}}$}. In the straight channel part pre-constriction, we only find a minor increase in the CFL between ROI1 and ROI2.1, indicative of reaching a saturation of the CFL thickness. Since the position of ROI2.1 corresponds to a distance of roughly \mbox{$120\times L/D_{\mathrm{h}}$}, we hypothesize that the experimental equilibrium CFL is reached before the RBCs enter the constriction. 

\begin{figure}[ht]
\centering
  \includegraphics[width=8.1cm]{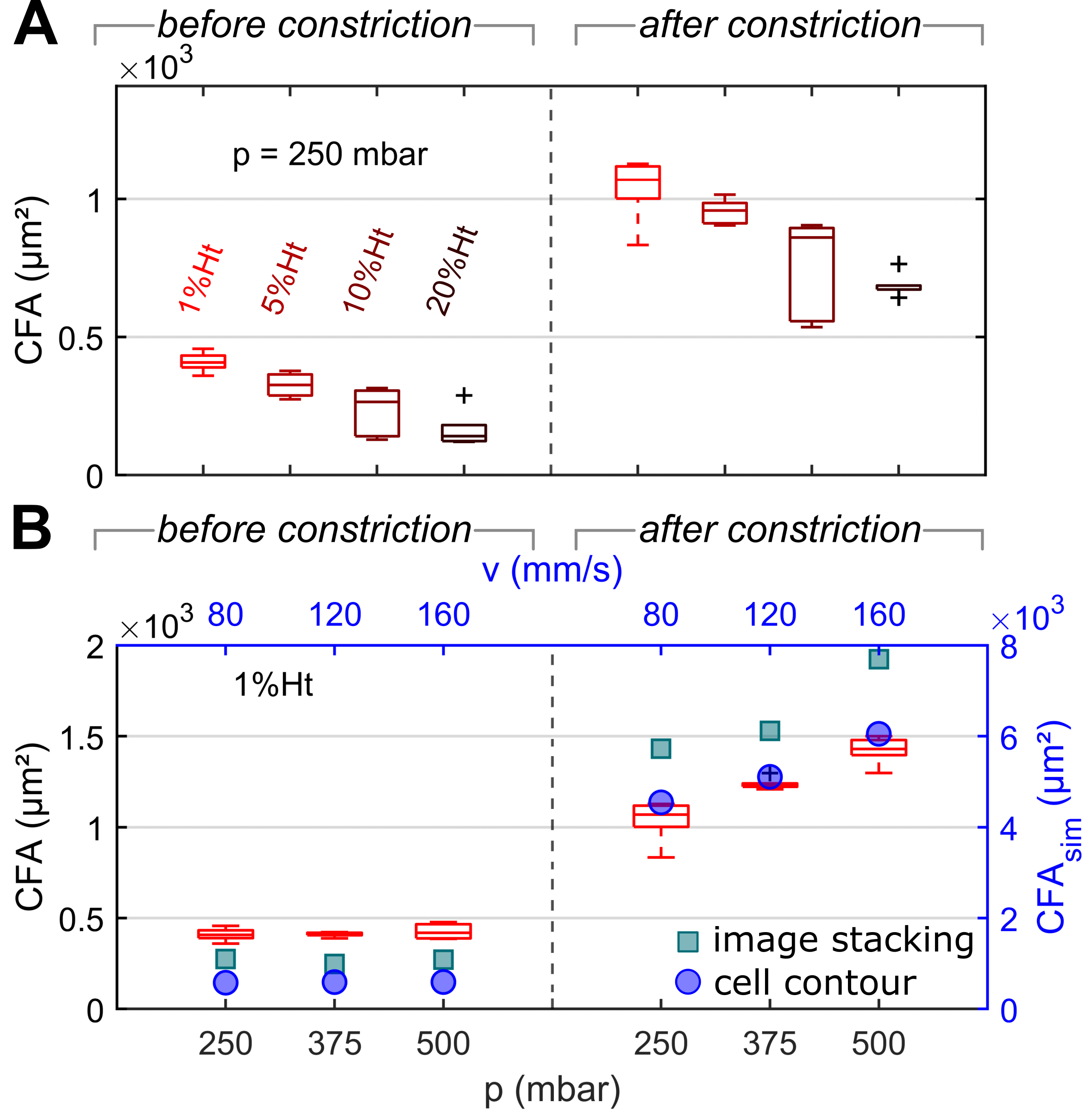}
  \caption{Effect of the RBC concentration and pressure drop on the CFA. (A) CFA for different RBC concentrations before and after the constriction in the experiment at a constant pressure drop of \mbox{$p=\unit[250]{mbar}$}. (B) CFA for different pressure drops for \mbox{$\unit[1]{\%Ht}$} RBC suspension before and after the constriction. Red box plots correspond to experimental data. Blue circles and teal squares correspond to numerical simulations, analyzed with the cell contour and image stacking method, respectively. The right $y$-axis \mbox{$\text{CFA}_{\text{sim}}$} in panel (B) corresponds to the numerical data.}
  \label{FIG_steadyCFAHt}
\end{figure}

In ROI2, we further investigate the effect of RBC concentration on the size of the CFA at the constriction. Figure~\ref{FIG_steadyCFAHt}A shows the size of the CFA in the experiments before and after the constriction as a function of the hematocrit at a fixed pressure drop of \mbox{$p=\unit[250]{mbar}$}. As exemplified in the snapshot of Fig.~\ref{FIG_steadyCFL}A, the CFAs before the constriction are smaller than after the constriction at a fixed pressure drop and hematocrit. Furthermore, the CFA size decreases with increasing RBC concentration, similar to the CFL thickness shown in Fig.~\ref{FIG_steadyCFLHt}. The ratio between the CFA size post/pre-constriction increases with the hematocrit from a factor of roughly three for \mbox{$\unit[1]{\%Ht}$} to a factor of four for the \mbox{$\unit[20]{\%Ht}$} RBC suspension, as shown in Fig.~S7B in the supplementary material. This increase is quantitively similar to the increase in the CFL ratio post/pre-constriction with the hematocrit. However, in contrast to the CFL ratio, the CFA ratio seems to increase with increasing pressure drop within the investigated range. This effect is also shown in Fig.~\ref{FIG_steadyCFAHt}B. While the CFA before the constriction does not significantly change with the applied pressure drop, the CFA post-constriction increases. A similar increase of the CFA post-constriction with the applied flow rate or pressure drop was reported in other constricted microchannels under steady flow conditions.\cite{Abay2020, RodriguezVillarreal2021} Here, we further find that the increase in CFA size with increasing pressure is more pronounced for small hematocrits, as shown in Fig.~S10 in the supplementary material. 

In our numerical simulations, we find a qualitatively similar behavior of the CFA pre- and post-constriction as a function of the flow rate, as shown by the circles and square symbols in Fig.~\ref{FIG_steadyCFAHt}B. Pre-constriction, the CFA is independent of the flow velocity \mbox{$v$}. However, a pronounced increase in the CFA size with \mbox{$v$} is observed post-constriction. Analog to the results reported in Fig.~\ref{FIG_steadyCFL}D, the image stacking method yields larger CFAs than the cell contour method. Note that the scaling of the CFA in the simulations in Fig.~\ref{FIG_steadyCFAHt}B is four times bigger than in the experiments. However, due to the increase in the constriction geometry size by a factor of two in every direction, we would expect much larger CFAs in the simulations. The absolute values of the CFA are also influenced by the shape of the contracting and expanding channel parts, which are different in the experiments and the numerical simulations (see Fig.~\ref{FIG_steadyCFL}A and B).   

\subsection{Time-dependent RBC flow at the constriction}
\subsubsection{Overview and definition of the CFLs and CFAs at the constriction during the sinusoidal pressure modulation}

\begin{figure}[ht]
\centering
  \includegraphics[width=8.1cm]{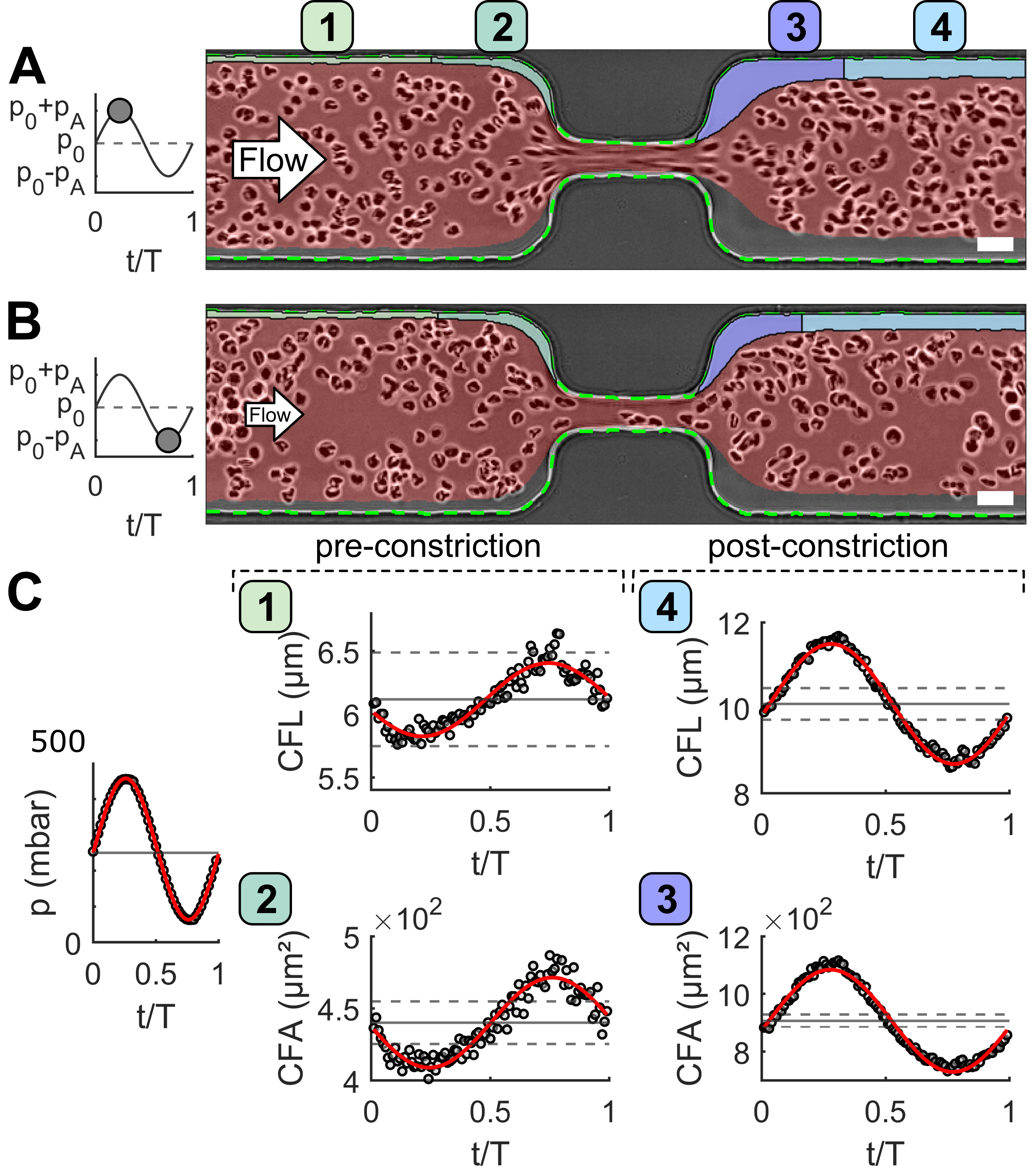}
  \caption{Determination of the CFL and the CFA during the time-dependent pressure modulation in the experiments. (A) and (B) show representative experimental snapshots of a \mbox{$\unit [5]{\%Ht}$} RBC suspension flow at the maximum (\mbox{$t/T=1/4$}) and minimum (\mbox{$t/T=3/4$}) pressure, respectively, using a sinusoidal modulation with a pressure offset of \mbox{$p_{0}=\unit[250]{mbar}$} and amplitude of \mbox{$p_\text{A}=\unit[200]{mbar}$}. Scale bars represent \mbox{$\unit[20]{\um}$}. Colored areas (1-4) in the snapshots in (A) and (B) indicate the CFL and CFA regions pre- and post-constriction. (C) Time-dependency of the CFL and CFA in the four colored regions in (A) and (B) at \mbox{$p_{0}=\unit[250]{mbar}$} and \mbox{$p_\text{A}=\unit[200]{mbar}$}. Gray symbols show the experimental data and red lines indicate sinusoidal fits. Horizontal full gray lines show the offset of the sinusoidal fits for each CFL or CFA. Dashed horizontal lines correspond to estimations of the experimental errors around the offset value for each parameter, as discussed in section~\ref{sec:CFLdetermination}.}
  \label{FIG_TimeDef}
\end{figure}

As shown in Fig.~\ref{FIG_steadyCFLHt}, the constriction drastically enhances the CFL under steady flow conditions. Here, we use a time-dependent flow to study the temporal dynamics of the cell-depleted regions focusing on the constriction region (ROI2 in Fig.~\ref{FIG_steadyDist}A). Therefore, we apply a sinusoidal pressure modulation \mbox{$p(t) = p_0+p_\text{A} \sin(\omega t)$} and use our image processing procedure to determine the position of the RBC core flow and the walls during the time-dependent flow conditions. Figure~\ref{FIG_TimeDef}A and B show experimental snapshots of the RBC core flow for different RBC concentrations during the maximum (\mbox{$p_0+p_\text{A}$}) and minimum (\mbox{$p_0-p_\text{A}$}) pressure in the cycle, respectively, employing a modulation with \mbox{$p_{0}=\unit[250]{mbar}$} and \mbox{$p_\text{A}=\unit[200]{mbar}$}. 

To quantify the size and temporal evolution of the CFAs, we define the CFL and the CFA in the constriction region according to the situation under steady flow conditions, as representatively illustrated for a \mbox{$\unit [5]{\%Ht}$} RBC suspension by the different colored areas for the top parts of the channels in Fig.~\ref{FIG_TimeDef}A and B. Note that under time-dependent flow conditions the transition points in $x$-direction between (1) and (2), and between (3) and (4) change along the flow direction during the pressure modulation, in contrast to measurements under steady flow. Hence, the size of the CFA does not only vary in $y$-direction but also in $x$-direction during the time-dependent flow modulation.

Figure~\ref{FIG_TimeDef}C shows representative data of the time-dependent behavior of the CFL and CFA for the four different regions using a \mbox{$\unit [5]{\%Ht}$} RBC suspension and a pressure modulation with \mbox{$p_{0}=\unit[250]{mbar}$} and \mbox{$p_{A}=\unit[200]{mbar}$}. The experimental data are fitted with sinusoidal functions using FFT, as indicated by the red lines in Fig.~\ref{FIG_TimeDef}D. Based on these fits, the offset, amplitude, and phase of the time-dependent signals are extracted and used for further analysis of the CFL and CFA dynamics at the constriction. 

For the CFL (regions 1 and 4 in Fig.~\ref{FIG_TimeDef}), we introduced a conservative error estimation based on the deformation of the channel wall, as discussed in section~\ref{sec:CFLdetermination}. Gray dashed lines in panels (1) and (4) of Fig.~\ref{FIG_TimeDef}C represent this error of \mbox{$\pm \unit[0.37]{\um}$} with respect to the mean CFL. While the time-dependent CFL post-constriction is larger than this error estimation, the CFL pre-constriction lies within these limits. To estimate an error in the CFA calculations due to a wall movement, we multiply the maximum observed wall deformation for the highest applied pressure drop, \textit{i.e.}, \mbox{$\unit[0.37]{\um}$}, with the maximum length of the CFA pre- and post-constriction. This serves as a conservative error estimation in \mbox{$\unit[]{\um^{2}}$} for the experimental CFA measurements and is plotted as gray dashed horizontal lines in panels (2) and (3) of Fig.~\ref{FIG_TimeDef}C. Similarly to the CFL, the time-dependent CFA signal is much larger than this error estimation post-constriction, while they are on the same order of magnitude before the constriction. 

In the following sections, we focus on the CFL offset \mbox{$\text{CFL}_0$} and amplitude \mbox{$\text{CFL}_\text{A}$} in section~\ref{CFL_dynamics}, and on the CFA offset \mbox{$\text{CFA}_0$} and amplitude \mbox{$\text{CFA}_\text{A}$} in section~\ref{CFA_dynamics}. Both error estimations are used in these sections for the CFL and CFA analysis. The phase behavior of the temporal CFL and CFA evolution with respect to the applied pressure signal is discussed in section~\ref{Phase_dynamics}.

\subsubsection{CFL dynamics at the constriction}\label{CFL_dynamics}

\begin{figure*}[ht]
\centering
  \includegraphics[width=\textwidth]{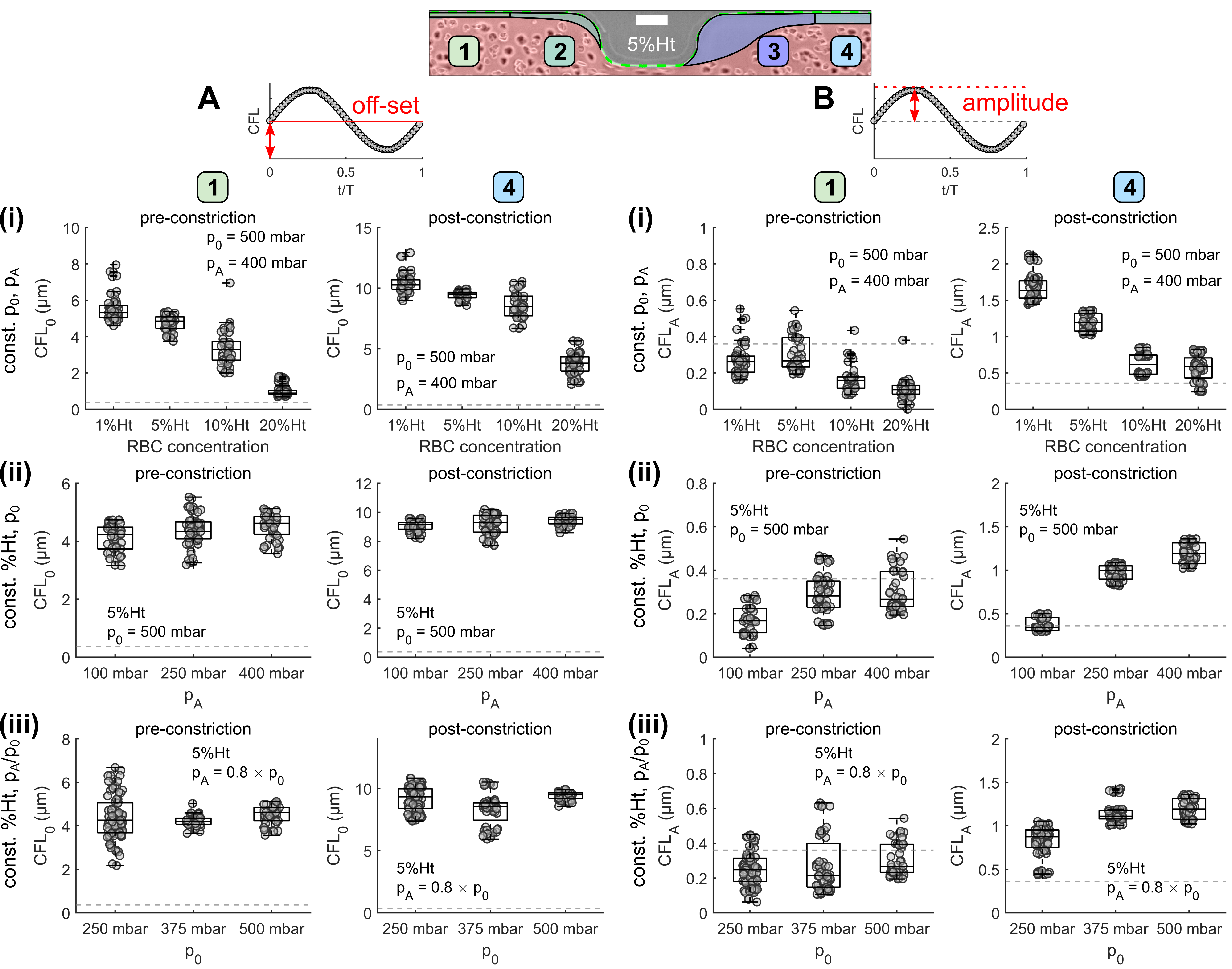}
  \caption{Offset \mbox{$\text{CFL}_0$} (A) and amplitude \mbox{$\text{CFL}_\text{A}$} (B) of the CFL under a sinusoidal pressure modulation at the constriction pre-constriction and post-constriction. Scale bars represent \mbox{$\unit[20]{\um}$}. The top panels in (A) and (B) show \mbox{$\text{CFL}_0$} and \mbox{$\text{CFL}_\text{A}$} for different RBC concentrations at \mbox{$p_{0}=\unit[500]{mbar}$} and \mbox{$p_\text{A}=\unit[400]{mbar}$}. The middle panels show the effect of the pressure amplitude \mbox{$p_\text{A}$} for a \mbox{$\unit [5]{\%Ht}$} RBC suspension at \mbox{$p_{0}=\unit[500]{mbar}$}. The bottom panels show \mbox{$\text{CFL}_0$} and \mbox{$\text{CFL}_\text{A}$} for a constant ratio of \mbox{$p_\text{A}/p_{0}=0.8$} for a \mbox{$\unit [5]{\%Ht}$} RBC suspension at different offset pressures \mbox{$p_{0}$}. Dashed horizontal lines indicate the error estimation for the highest applied pressure drop.}
  \label{FIG_CFL}
\end{figure*}

Figure~\ref{FIG_CFL} shows the dependency of the CFL offset \mbox{$\text{CFL}_0$} (A) and amplitude \mbox{$\text{CFL}_\text{A}$} (B) on the RBC concentration and the parameters of the sinusoidal pressure modulation (\mbox{$p_{0}$} and \mbox{$p_\text{A}$}). The left and right panels in Fig.~\ref{FIG_CFL}A and B correspond to the pre-constriction (1) and post-constriction (4) regions, respectively. 

The top row in Fig.~\ref{FIG_CFL} shows the effect of an increasing RBC concentration on \mbox{$\text{CFL}_0$} and \mbox{$\text{CFL}_\text{A}$} at a fixed \mbox{$p_{0}=\unit[500]{mbar}$} and \mbox{$p_\text{A}=\unit[400]{mbar}$}. Similar to the observations under steady flow conditions, the CFL offset \mbox{$\text{CFL}_0$} decreases with increasing RBC concentration both pre-constriction and post-constriction. The magnitude of the \mbox{$\text{CFL}_0$} is in good agreement with the results under steady flow conditions, shown in Fig.~\ref{FIG_steadyCFLHt}A. The ratio of the \mbox{$\text{CFL}_0$} post/pre-constriction is independent \mbox{$p_\text{A}$}, as shown in Fig.~S11 in the supplementary material. However, it increases with increasing RBC concentration from a factor of two for \mbox{$\unit [5]{\%Ht}$} to a factor between \mbox{$3-4$} for the \mbox{$\unit [20]{\%Ht}$} suspension. This dependency is similar to the observations under steady flow conditions (see Fig.~S7A). We further observe merely a slight increase of the ratio with \mbox{$p_{0}$}. Qualitatively similar dependencies of the CFL ratio post/pre-constriction on the hematocrit and the applied flow rate or \mbox{$p_{0}$} have been reported under steady flow conditions.\cite{Faivre2006, Yaginuma2013} However, these observations were only reported for low Re and for low RBC concentrations \mbox{$\leq \unit [2.6]{\%Ht}$}.

While the amplitude of the CFL oscillations \mbox{$\text{CFL}_\text{A}$} pre-constriction is very small and within the established error limit of the CFL oscillations (dashed vertical lines in Fig.~\ref{FIG_CFL}B(i)), we find a decrease in \mbox{$\text{CFL}_\text{A}$} with increasing hematocrit post-constriction. The ratio of the \mbox{$\text{CFL}_\text{A}$} post/pre-constriction is roughly five and does not depend on the RBC concentration, as shown in Fig.~S11B. 

Keeping the RBC concentration and \mbox{$p_{0}$} constant and changing the amplitude \mbox{$p_\text{A}$} of the pressure signal does not result in a significant variation of \mbox{$\text{CFL}_0$} pre-constriction and post-constriction, as shown in Fig.~\ref{FIG_CFL}A(ii) and Fig.~S12A in the supplementary material. However, increasing \mbox{$p_\text{A}$} leads to an increase of the CFL amplitude \mbox{$\text{CFL}_\text{A}$} for all hematocrits pre-constriction and post-constriction (Fig.~\ref{FIG_CFL}B(ii) and Fig.~S12B). The ratio of the \mbox{$\text{CFL}_\text{A}$} post/pre-constriction increases with increasing \mbox{$p_\text{A}$} for RBC concentrations below \mbox{$\unit [10]{\%Ht}$} (Fig.~S11D). Similar to the results under steady flow conditions, the CFL offset \mbox{$\text{CFL}_0$}, as well as the amplitude \mbox{$\text{CFL}_\text{A}$} under sinusoidal driving, is not systematically influenced by the applied pressure offset \mbox{$p_{0}$} at a fixed \mbox{$p_\text{A}$} within the investigated range, as shown in Fig.~\ref{FIG_CFL}(iii). 

The results presented in Fig.~\ref{FIG_CFL} demonstrate that the CFL dynamics under time-dependent flow conditions are foremost influenced by the RBC concentration and the amplitude of the pressure modulation. Further investigated \mbox{$p_{0}$}-\mbox{$p_\text{A}$}-combinations and hematocrits are shown in Fig.~S12 and exhibit the same behavior. The amplitude of the CFL oscillations \mbox{$\text{CFL}_\text{A}$} pre-constriction are within the experimental detection limit of the used setup. However, the oscillations post-contraction can be reasonably resolved, and a pronounced impact of the pressure amplitude and the hematocrit on \mbox{$\text{CFL}_\text{A}$} can be established. The observed offset values \mbox{$\text{CFL}_0$} and their dependencies on the flow parameters shown in Fig.~\ref{FIG_CFL}A are in agreement with the results of Fig.~\ref{FIG_steadyCFLHt}A and similar experimental studies under steady flow conditions.\cite{Fujiwara2009, Bento2019, Abay2020} Notably, Faivre~\etal~\cite{Faivre2006} found that the CFL in their constricted channel increased with increasing hematocrit, opposite to the results presented in this study. However, the authors merely covered a small RBC concentration range of \mbox{$\unit [0.1-2.6]{\%Ht}$}. In higher concentrated RBC suspension, the hydrodynamic interactions between RBCs become progressively important.  Therefore, we believe that the deviation between our observation and the work of Faivre~\etal~\cite{Faivre2006} regarding the influence of the hematocrit on the CFL stems from the difference in the investigated RBC regimes. Furthermore, we note that the changes in the magnitude of the CFL induced by the constriction are qualitatively similar to those caused by microchannel networks.\cite{Bento2019}

\subsubsection{CFA dynamics at the constriction}\label{CFA_dynamics}

\begin{figure*}[ht]
\centering
  \includegraphics[width=\textwidth]{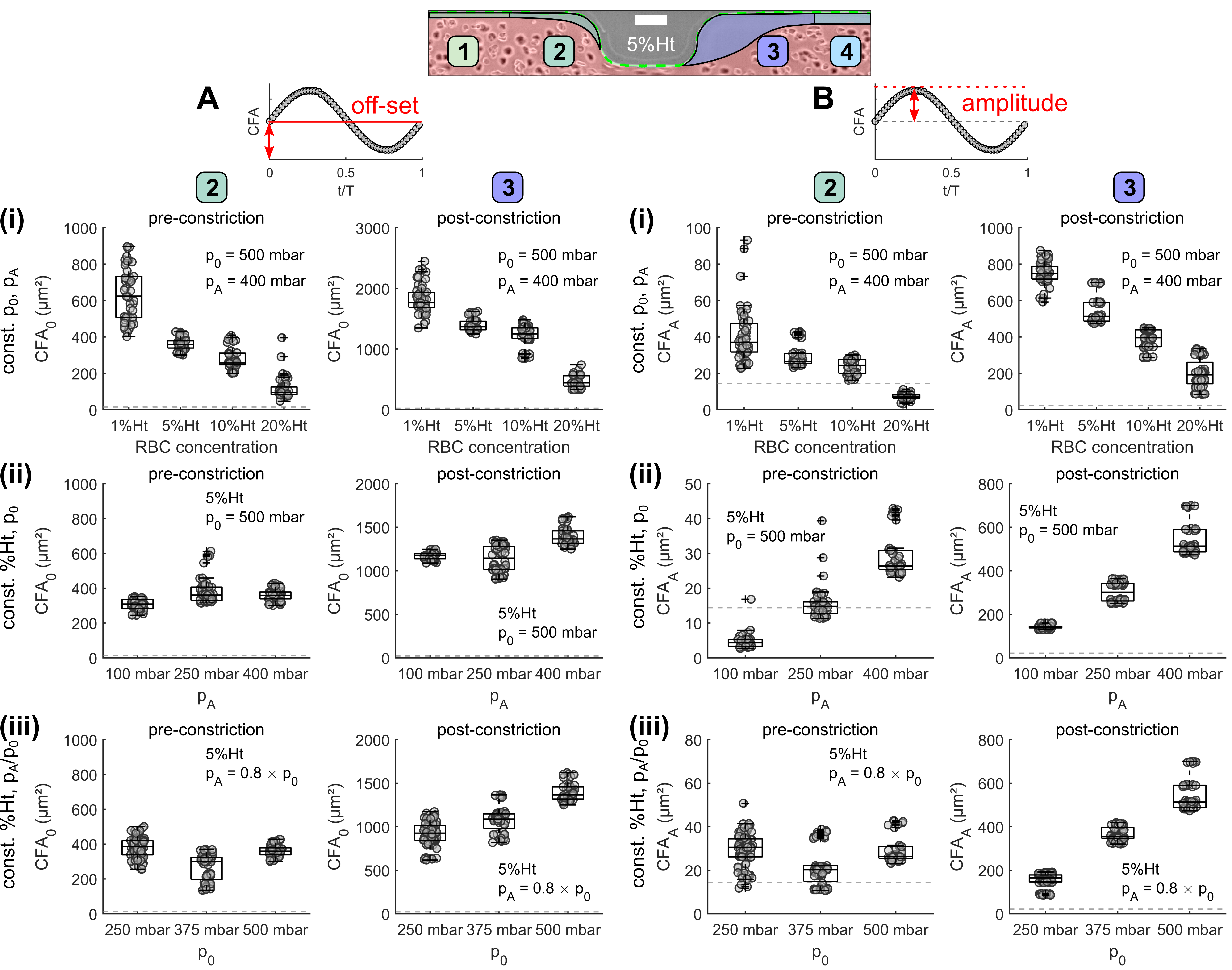}
  \caption{Offset \mbox{$\text{CFA}_0$} (A) and amplitude \mbox{$\text{CFA}_\text{A}$} (B) of the CFA under a sinusoidal pressure modulation at the constriction pre-constriction and post-constriction. Scale bars represent \mbox{$\unit[20]{\um}$}. The top panels in (A) and (B) show \mbox{$\text{CFA}_0$} and \mbox{$\text{CFA}_\text{A}$} for different RBC concentrations at  \mbox{$p_{0}=\unit[500]{mbar}$} and \mbox{$p_\text{A}=\unit[400]{mbar}$}. The middle panels show the effect of the pressure amplitude \mbox{$p_\text{A}$} for a \mbox{$\unit [5]{\%Ht}$} RBC suspension at \mbox{$p_{0}=\unit[500]{mbar}$}. The bottom panels show \mbox{$\text{CFA}_0$} and \mbox{$\text{CFA}_\text{A}$} for a constant ratio of \mbox{$p_\text{A}/p_{0}=0.8$} for a \mbox{$\unit [5]{\%Ht}$} RBC suspension at different offset pressures \mbox{$p_{0}$}. Dashed horizontal lines indicate the error estimation for the highest applied pressure drop.}
  \label{FIG_CFA}
\end{figure*}

Similar to the results of the CFL analysis, Fig.~\ref{FIG_CFA} shows the dependency of the CFA offset \mbox{$\text{CFA}_0$} (A) and amplitude \mbox{$\text{CFA}_\text{A}$} (B) on the RBC concentration and the parameters of the sinusoidal pressure modulation (\mbox{$p_{0}$} and \mbox{$p_\text{A}$}). The left and right panels in Fig.~\ref{FIG_CFA}A and B correspond to the pre-constriction (2) and post-constriction (3) regions, respectively. 

The size of the CFAs and hence \mbox{$\text{CFA}_0$} decrease both pre-constriction and post-constrictions with increasing RBC concentration, as shown in Fig.~\ref{FIG_CFA}A(i) at a fixed \mbox{$p_{0}=\unit[500]{mbar}$} and \mbox{$p_\text{A}=\unit[400]{mbar}$}. Related to the observation for the CFL, the ratio between the \mbox{$\text{CFA}_0$} post/pre-constriction does not depend on \mbox{$p_\text{A}$} but increases with increasing hematocrit, as shown in Fig.~S13A and C in the supplementary material. It increases by a factor of roughly two between \mbox{$\unit [1]{\%Ht}$} and \mbox{$\unit [20]{\%Ht}$}. In contrast to the CFL ratio, the CFA ratio exhibits a more pronounced dependency on \mbox{$p_0$} and increases with the applied pressure offset. This is similar to the observation of the CFA ratio under steady flow conditions, shown in Fig.~S7B. The CFA amplitude \mbox{$\text{CFA}_\text{A}$} also decreases with increasing RBC concentration (Fig.~\ref{FIG_CFA}B(i)). However, the \mbox{$\text{CFA}_\text{A}$} ratio post/pre-constriction is more than one order of magnitude larger than the \mbox{$\text{CFA}_0$} ratio, independent of the RBC concentration and \mbox{$p_\text{A}$}, shown in Fig.~S13B and D.

The CFA offset \mbox{$\text{CFA}_0$} is not significantly influenced by the amplitude of the pressure modulation (Fig.~\ref{FIG_CFA}A(ii)). In contrast, we observe a drastic increase of the CFA amplitude \mbox{$\text{CFA}_\text{A}$} with \mbox{$p_\text{A}$}, as demonstrated for \mbox{$\unit [5]{\%Ht}$} in Fig.~\ref{FIG_CFA}B(ii). While \mbox{$\text{CFA}_0$} and \mbox{$\text{CFA}_\text{A}$} seems to be independent of the applied pressure offset \mbox{$p_0$} pre-constriction, we observe an increase of both quantities with \mbox{$p_0$} after the constriction, as shown in Fig.~\ref{FIG_CFA}(iii).  
We observe the same qualitative behavior regarding the CFA dynamics for all other investigated \mbox{$p_{0}$}-\mbox{$p_\text{A}$}-combinations and RBC concentrations, as described in Fig.~S14 in the supplementary material.

Similar to the CFL, the CFA dynamics, characterized by \mbox{$\text{CFA}_0$} and \mbox{$\text{CFA}_\text{A}$} in Fig.~\ref{FIG_CFA}, are primarily influenced by the RBC concentration. The most striking feature is the strong increase of the amplitude of the CFA oscillations \mbox{$\text{CFA}_\text{A}$} with the amplitude of the sinusoidal pressure modulation.

So far, most experimental studies have focused on the downstream vortices and cell-depleted zones under steady flow conditions since they present a favorable location for plasma extraction.\cite{KersaudyKerhoas2010a, Sollier2010, Marchalot2014a, Tripathi2015} A similar increase of the CFA post-constriction with the applied flow rate or pressure drop was recently reported, yet under steady flow conditions.\cite{RodriguezVillarreal2021} In this study, we use a symmetric contraction-expansion geometry with curved walls that gradually narrow down and increase. Note that sharp and cornered expansions can lead to the formation of lip vortices close to the reentrant corners post-constriction that have been observed in RBC suspensions and blood analog fluid.\cite{RodriguezVillarreal2021, Sousa2011a, Sousa2011b}  With increasing inertia, these vortices grow in size in the downstream direction until the vortices and CFA stretch toward the channel side wall,\cite{Abay2020} similar to our observations.

\begin{figure}[ht]
\centering
  \includegraphics[width=8.3cm]{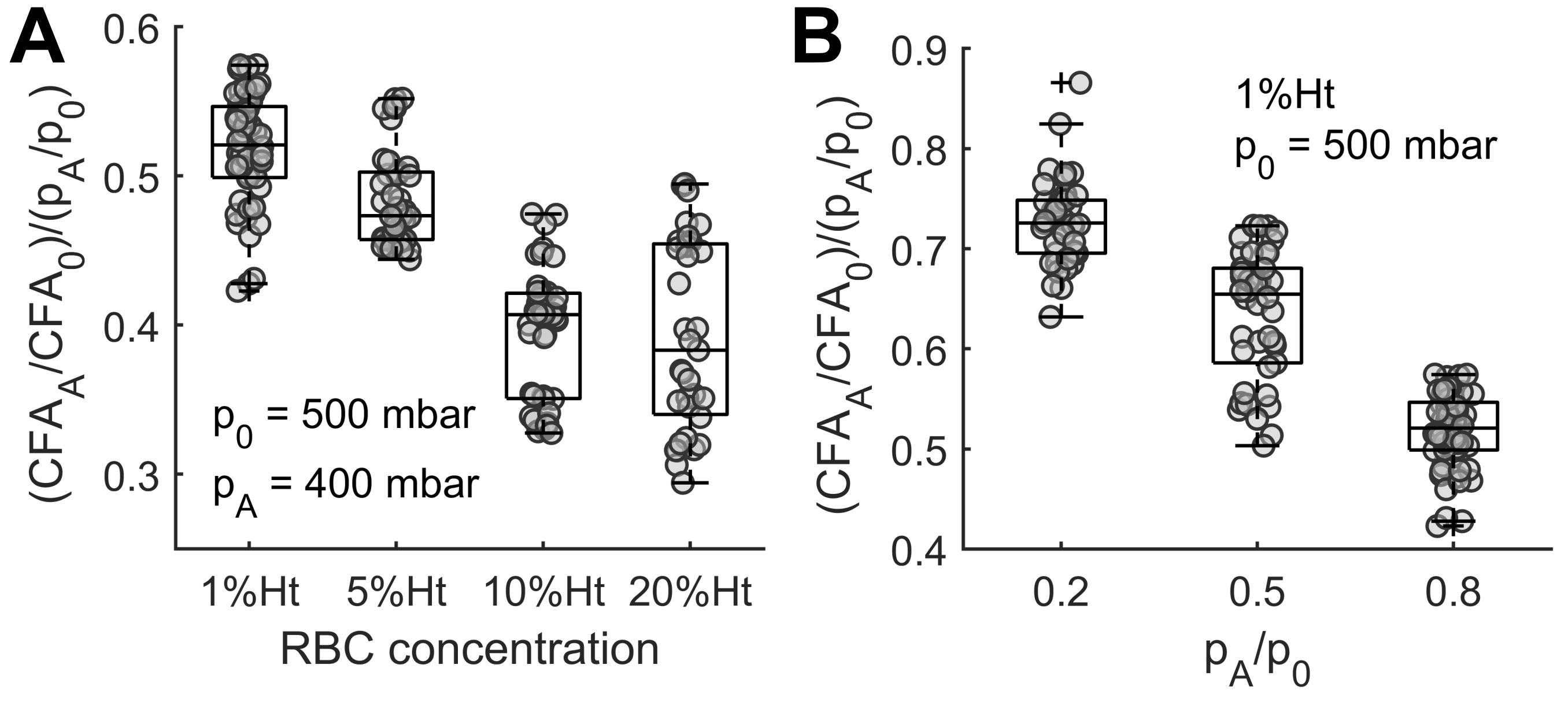}
  \caption{Ratio between the relative amplitudes of the CFA and pressure oscillations \mbox{$(\text{CFA}_\text{A}/\text{CFA}_0)/(p_\text{A}/p_{0})$} 
  post-constriction. (A) Ratio as a function of the RBC concentration at \mbox{$p_{0}=\unit[500]{mbar}$} and \mbox{$p_\text{A}=\unit[400]{mbar}$}. (B) Ratio as a function of \mbox{$p_\text{A}/p_{0}$} at \mbox{$p_{0}=\unit[500]{mbar}$} and \mbox{$\unit [1]{\%Ht}$}.}
  \label{FIG_Aratio}
\end{figure}

Besides the observations regarding the static CFA size, our investigations under time-dependent flow conditions allow us to relate the oscillations of the CFA to the applied pressure signal. Therefore, we compare the relative amplitude of the CFA oscillations \mbox{$(\text{CFA}_\text{A}/\text{CFA}_0)$} to the relative amplitude of the pressure modulation \mbox{$(p_\text{A}/p_{0})$}.  Because the \mbox{$\text{CFA}_\text{A}$} pre-constriction is larger than our error estimation only at large pressure amplitudes, \textit{e.g.}, at \mbox{$p_{0}=\unit[500]{mbar}$} and \mbox{$p_\text{A}=\unit[400]{mbar}$} shown in Fig.\ref{FIG_CFA}B(ii), we calculate the ratio between both parameters \mbox{$(\text{CFA}_\text{A}/\text{CFA}_0)/(p_\text{A}/p_{0})$} only for the post-constriction region. Post-constriction, we find that the relative amplitude ratio \mbox{$(\text{CFA}_\text{A}/\text{CFA}_0)/(p_\text{A}/p_{0})$}  decrease with increasing hematocrit, as shown in Fig.~\ref{FIG_Aratio}A. Hence, we hypothesize that the presence of the deformable RBCs in the suspending medium results in a dampening of the pronounced CFA oscillations. 

Further, the ratio increases with decreasing relative pressure amplitude \mbox{$(p_\text{A}/p_{0})$}, as shown in Fig.~\ref{FIG_Aratio}B for \mbox{$\leq\unit[1]{\%Ht}$}. Consequently, we find that the relative amplitude ratio \mbox{$(\text{CFA}_\text{A}/\text{CFA}_0)/(p_\text{A}/p_{0})$} approaches unity post-constriction for the lowest investigated RBC concentration of \mbox{$\unit[1]{\%Ht}$} and the smallest relative pressure amplitude of \mbox{$p_\text{A}/p_{0}=0.2$}. This observation could indicate an influence of transient inertia effects at \mbox{$\text{Re}>1$}, although \mbox{$\text{Wo}\ll1$}. As time-dependent flows of RBCs or other complex particulate systems are sparsely investigated today, such phenomena remain poorly understood, especially since the generation of flow oscillations in microfluidic at high frequencies with commercial pressure controllers remains challenging and can lead to significant deviations from the desired waveform.\cite{Recktenwald2021b} However, recent advantages in the generation of dynamic vortices in microfluidic systems using tube oscillations\cite{Thurgood2022a, Thurgood2022b} could be used in future investigations in combination with our CFL detection method to examine such transient inertia effects at elevated frequencies.

\subsubsection{Phase shift of the CFA before and after the constriction}\label{Phase_dynamics}

\begin{figure}[ht]
\centering
  \includegraphics[width=8.3cm]{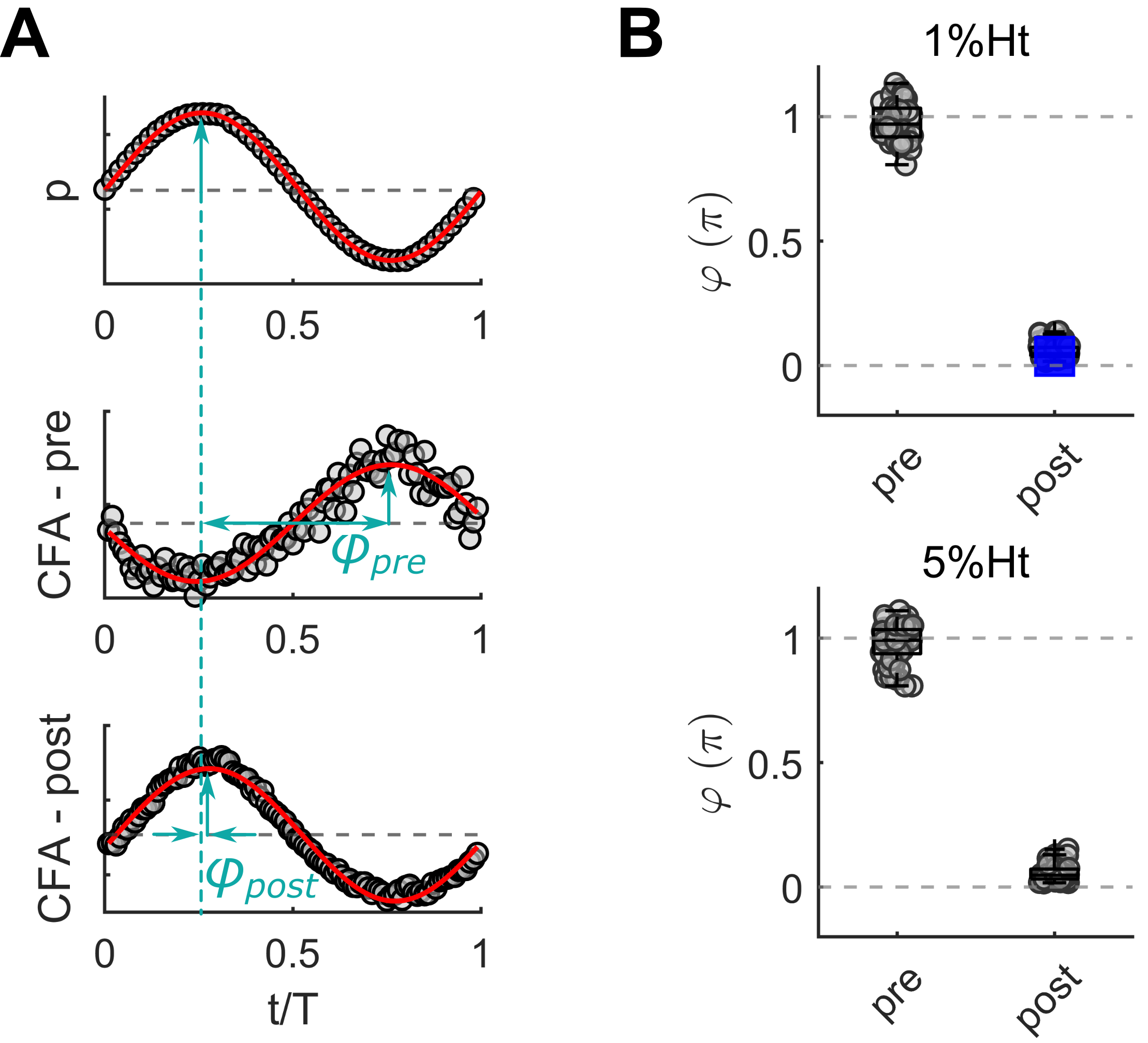}
  \caption{Phase shift of the CFA. (A) Time-dependent pressure (top) and representative evaluations of the CFA pre-constriction (middle) and CFA post-constriction (bottom) for pressure modulation are shown in (A). Gray symbols show the experimental data and red lines indicate sinusoidal fits. Phase differences pre- \mbox{$\varphi_\text{pre}$} and post-constriction \mbox{$\varphi_\text{post}$} are schematically depicted as teal arrows. (B) Phase difference \mbox{$\varphi$} of the CFA pre- and post-constriction for \mbox{$\unit [1]{\%Ht}$}  and \mbox{$\unit[5]{\%Ht}$} RBC suspensions at \mbox{$p_{0}=\unit[500]{mbar}$} and \mbox{$p_\text{A}=\unit[400]{mbar}$}. The blue symbol in (B) corresponds to the numerical simulation at \mbox{$\unit [1]{\%Ht}$}}.
  \label{FIG_Shift}
\end{figure}

In section~\ref{CFL_dynamics} and \ref{CFA_dynamics}, we characterized the offset and amplitude of the time-dependent CFL and CFA during the sinusoidal pressure modulation at the constriction. Here, we examine the phase of the temporal CFA with respect to the applied pressure signal. Since the time-dependent CFL signal is within the error estimation pre-constriction, and because the CFL and CFA decrease with increasing hematocrit and decreasing pressure, we limit our phase shift analysis to the CFA data at the highest pressure modulations (\mbox{$p_{0}=\unit[500]{mbar}$} and \mbox{$p_\text{A}=\unit[400]{mbar}$}) and the smallest RBC concentrations (\mbox{$\unit [1]{\%Ht}$}  and \mbox{$\unit[5]{\%Ht}$}). Figure~\ref{FIG_Shift}A schematically shows a time-dependent pressure signal (top) and representative evaluations of the CFA signals pre-constriction (middle) and post-constriction (bottom). Based on the sinusoidal fits of the signal, indicated by the red lines in Fig.~\ref{FIG_Shift}A, we determine the phase shift \mbox{$\varphi=\phi_p-\phi_i$}, with the phase of the pressure modulation \mbox{$\phi_p$} and the phases of the CFA signals \mbox{$\phi_i$} pre- or post-constriction. Teal arrows in Fig.~\ref{FIG_Shift}A exemplify the phase shift of the CFA pre-constriction \mbox{$\varphi_\text{pre}$} and post-constriction \mbox{$\varphi_\text{post}$}.

Figure~\ref{FIG_Shift}B summarized the analysis of the phase shift pre- and post-constriction of the investigated \mbox{$\unit [1]{\%Ht}$}  and \mbox{$\unit[5]{\%Ht}$} RBC concentrations. While the temporal CFA evolution post-constriction is in-phase with the applied pressure modulation, we observe a phase inversion of the CFA pre-constriction. Here, the CFA signal before the constriction is out-of-phase with a \mbox{$\unit[180]{^\circ}$} phase shift. 

\begin{figure}[ht]
\centering
  \includegraphics[width=8.3cm]{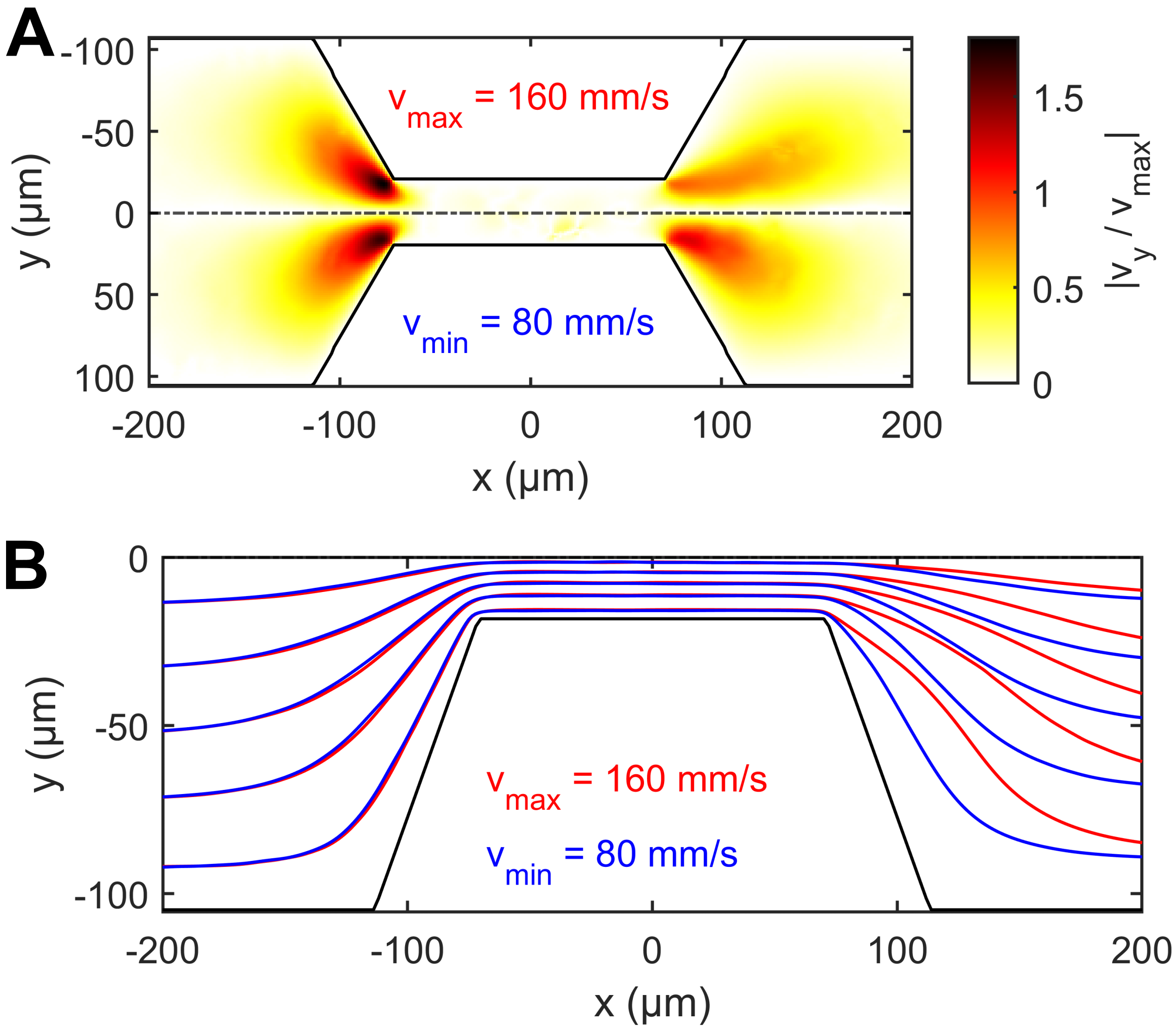}
  \caption{Results of the numerical simulation during time-dependent flow at the maximum \mbox{$v_{\text{max}}$} and minimum \mbox{$v_{\text{min}}$} velocities during the flow modulation. (A) Absolute values of the normalized velocity component \mbox{$v_y$} in lateral $y$-direction at the constriction. The top half of panel (A) shows the top half of the constriction at \mbox{$v_{\text{max}}$}, while the bottom half of the panel shows the bottom half of the constriction at \mbox{$v_{\text{min}}$}. (B) Streamlines at \mbox{$v_{\text{max}}$} and \mbox{$v_{\text{min}}$} at the bottom half of the constriction. Each streamline pair starts at the same $y$-position at the beginning of the channel.}
  \label{FIG_SIMvy}
\end{figure}

To understand this phase shift in the experiments, we perform numerical simulations under a time-dependent driving of the flow. While we find an in-phase behavior of the CFA post-constriction in the simulations (see the blue symbol in Fig.~\ref{FIG_Shift}B at \mbox{$\unit [1]{\%Ht}$}), in agreement with our microfluidic experiments, the CFA pre-constriction scattered broadly over time and did not allow us to fit the signal with a sine function, as shown in Fig.~S15 in the supplementary material. In addition, we extracted the velocity field of the fluid during the sinusoidal flow modulation in our 3D numerical simulations. Pre-constriction, there is a flow towards the centerline, while post-constriction the flow has a $y$-component directed towards the channel walls. Figure~\ref{FIG_SIMvy}A shows the normalized velocity component \mbox{$v_y$} along the constriction for a horizontal channel slice at \mbox{$z=H/2$}.  The flow field is symmetric with respect to the channel centerline at \mbox{$y = 0$}, therefore, in the top half of panel (A) we depict the flow in the top half of the constriction at \mbox{$v_{\text{max}}$} (the maximum flow velocity during the sinusoidal oscillation) and the bottom half of panel (A) shows the flow in the bottom half of the constriction at \mbox{$v_{\text{min}}$} (the minimum during sinusoidal oscillation). In 
Fig.~\ref{FIG_SIMvy}A, we can see that the lateral flow components post-constriction differ visibly. As the velocity increases during the
cycle to \mbox{$v_{\text{max}}$}, the high-\mbox{$v_y$} region is extended further downstream while at \mbox{$v_{\text{min}}$}, this region is extended more towards the channel wall.
Furthermore, the maximal strength of this normalized lateral flow is larger for \mbox{$v_{\text{min}}$}.

Additionally, we plot corresponding streamlines at \mbox{$v_{\text{max}}$} and \mbox{$v_{\text{min}}$} in Fig.~\ref{FIG_SIMvy}B, both in the lower channel half and again at \mbox{$z=H/2$}. Each streamline pair starts at the same $y$-position at the beginning of the channel. The streamlines at \mbox{$v_{\text{min}}$} post-constriction are bent closer to the channel walls. Hence, there is a pronounced difference between the two velocities in Fig.~\ref{FIG_SIMvy}B. Pre-constriction, the difference between \mbox{$v_{\text{min}}$} and \mbox{$v_{\text{max}}$} is less pronounced but still visible in the streamline plot, the orientation of the streamlines is
reversed, \textit{i.e.} the streamlines at \mbox{$v_{\text{max}}$} run slightly closer
to the channel walls. 

At higher velocities differences (\mbox{$\Delta v = v_{\text{max}} - v_{\text{min}}$}), hence higher \mbox{$\Delta\text{Re}$}, these differences would increase. Note that the numerical simulations are performed at velocities that correspond to the lowest pressure drop combinations in the experiments (\mbox{$p_{0}=\unit[250]{mbar}$} and \mbox{$p_\text{A}=\unit[200]{mbar}$}), where the phase shift of the CFA pre-constriction is within the experimental error. For the higher pressure drops reported in Fig.~\ref{FIG_Shift}, we hypothesize that the RBCs follow streamlines that are bent against the narrowing constriction wall pre-constriction when the pressure increases during the cycle in the experiment. Hence, the size of the CFA before the constriction decreases. During the receding phase of the pressure cycle, these areas show a minor relaxation and expand again, resulting in an increase in the CFA size pre-constriction. Therefore, we observe a \mbox{$\unit[180]{^\circ}$} phase shift for the CFA dynamics before the construction in the experiments. In contrast, the CFA post-constriction increases in size during the high-pressure phase of the pressure cycle. Here, RBCs are accelerated rapidly when passing through the narrow throat of the constriction and an enhanced jet flow is observed with vortices in the corners downstream of the expansion.\cite{Abay2020} This results in an increase in the CFA size. During the receding phase of the pressure cycle, the magnitude of the vortices and the jet flow decreases, allowing the RBCs to flow closer to the channel walls, which reduces the CFA size post-constriction. Thus, the CFA directly after the constriction is in-phase with the applied pressure signal.

To support this experimental hypothesis, numerical simulations would have to be performed at higher flow rates and with higher velocity amplitudes. However, with the current simulation setup, we are not able to achieve velocities above \mbox{$\approx\unit[170]{mm/s}$} (\mbox{$\text{Re}\approx10$}). Increasing the velocity beyond this value results in instabilities of the discretized RBC membrane in the high shear rate zones in the constriction. 

In the context of time-dependent microfluidic flows, the observed phase shift phenomenon at the constriction in the experiments could be examined for other complex fluids such as blood analog solutions and RBC suspension in plasma. Due to their non-Newtonian flow behavior, characterized by pronounced shear thinning in contrast to the fluids with a constant viscosity used in this study (see Fig.~S3), and with a finite relaxation time, such fluids generate both upstream and downstream vortices and recirculation areas.\cite{Sousa2011a, Sousa2011b, Brust2013} In combination with microfluidic designs that mimic complex in-vivo situations, such as interactions of vortical flows with RBCs in venous valves\cite{Sanchez2022}, our CFL detection method could be readily applied to advance the field of hemorheology under physiological conditions.

\section{Conclusions}
Studying the microfluidic flow behavior of RBCs advances our knowledge of blood flow in-vivo and is crucial to understand pathological changes in the circulatory system. Additionally, microfluidic investigations are the basis for lab-on-a-chip designs for diagnostic and biomedical applications. In such devices, as well as in the vessels of the cardiovascular system, constrictions and sudden changes in the vessel cross-section dramatically impact the spatiotemporal cell organization and RBC flow. 

In this study, we performed microfluidic experiments to understand the RBC flow behavior and the CFL phenomenon in a constricted microchannel covering a broad hematocrit range up to \mbox{$\unit [20]{\%Ht}$}. Besides investigations under steady flow, we examine the RBC and CFL dynamics at the constriction under time-dependent flow conditions, which have recently received increasing attention in the microfluidic community due to their potential to enhance various microscale operations and as biomimicry for biological processes.

To examine the CFL dynamics in a non-stationary flow, we first developed a post-processing routine that enables us to detect the time-dependent RBC core flow and the positions of the channel borders based on image sequences from optical microscopy. In the present work, we employ sinusoidal pressure modulations to generate a time-dependent RBC flow. However, the proposed method can be used for any arbitrary periodic waveform, such as blood pressure waveforms, in both pressure-driven flows or by applying a time-dependent volume flow rate. Employing this method, we show how the temporal dynamics of the CFL and the CFA before and after the constriction are influenced by the hematocrit and the offset and amplitude of the applied flow modulation. 

Our findings highlight the importance of understanding the phenomena of particle-laden flows under time-dependent flow conditions. Especially the amplitude of the CFA and the phase shift between the CFA pre-constriction and the pressure signal, in contrast to the in-phase behavior of the CFA post-constriction, provide an interesting possibility for novel lab-on-a-chip plasma separation devices. Furthermore, this phenomenon could be studied in the context of trapping rigid particles or RBCs with impaired deformability, which are found in multiple diseases, such as malaria, diabetes, sickle cell disease, or acanthocytosis,\cite{Stuart1990, Symeonidis2001, Dondorp2002, Mannino2012, Reichel2022} within the vortex and recirculation regions of the constriction.\cite{Abay2020} Moreover, our method can be readily applied to investigate time-dependent RBC flow in other complex geometries, such as microchannel networks that were shown to affect the CFL in successive vessels.\cite{Bento2018, Balogh2019, Bento2019, Li2022a}

Further, our experimental results enable us to resolve the time-dependent flow field at the constriction and provide a basis for future numerical simulations of flow phenomena in unsteady flows. Our initial simulations under steady flow conditions are qualitatively in good agreement with the experiments regarding the RBC distribution as well as the CFL and CFA behavior. Although our numerical results are limited to \mbox{$\text{Re}<10$}, we observed differences in the  time-dependent flow behavior pre- and post-constriction. Nevertheless, future robust simulations at elevated \mbox{$\text{Re}>10$} should be performed to test the experimental hypotheses. 

We believe that the presented work can be used as a methodological framework for future experimental and numerical studies in the emerging field of unsteady microscale flows,\cite{Dincau2020} which many researchers have begun to incorporate in their experiments on droplet generation, mixing, particle-cell separation, clog mitigation, and biomimicry in physiological studies.

\begin{acknowledgments}
This work was funded by the Deutsche Forschungsgemeinschaft (DFG, German
Research Foundation) – project number 349558021 (RE~5025/1-2, WA 1336/13-1, and GE 2214/2-1). K.G. thanks the `Studienstiftung des Deutschen Volkes' for financial support. 
\end{acknowledgments}

\bibliography{references}

\end{document}